\documentclass{tlp}
\usepackage{url}
\usepackage{mdv_macros}
\usepackage[pdftex]{graphicx}
\usepackage{booktabs}
\submitted{15 June 2009}
\revised{15 Oct 2009}

\title[Music Composition using ASP]{Automatic Music Composition using Answer Set Programming}

\author[G. Boenn et al.]{GEORG BOENN\\
Cardiff School of Creative \& Cultural Industries,\\
University of Glamorgan, \\
Pontypridd, CF37 1DL, UK\\
\email{gboenn@glam.ac.uk}
\and MARTIN BRAIN, MARINA DE VOS and JOHN FFITCH\\
Department of Computer Science, \\
University of Bath, \\
Bath, BA2 7AY, UK\\
\email{\{mjb,mdv,jpff\}@cs.bath.ac.uk}
}

\newcommand{\systemname}[0]{\textsc{Anton}}
\newcommand{\systemnamev}[1]{\textsc{Anton} v.#1}
\newcommand{\AnsProlog}[0]{\ensuremath{AnsProlog} }
\newcommand{\AnsPrologs}[0]{\ensuremath{AnsProlog}}
\newcommand{\filename}[1]{\texttt{#1}}
\newtheorem{definition}{Definition}

\newcommand{\prog}[1]{\textsc{#1}}

\begin{document}

\newcommand{\note}{\begin{quote}\bf\end{quote}}
\pagerange{\pageref{firstpage}--\pageref{lastpage}}
\volume{\textbf{??} (?):}
\jdate{??}
\setcounter{page}{1}
\pubyear{??}

\maketitle

\label{firstpage}

\begin{abstract}
Music composition used to be a pen and paper activity. These 
these days music is often composed with the aid of computer software, even
to the point where the computer compose parts of 
the score autonomously.

The composition of most styles of music is governed by rules. 
We show that by approaching the automation, analysis and verification of
composition as a knowledge representation task and formalising these
rules in a suitable logical language,  
powerful and expressive intelligent composition tools can be easily built.  

This application paper describes the use of answer set
programming to construct an automated system, named \systemname, that
can compose melodic, harmonic and rhythmic music, diagnose errors in human
compositions and serve as a computer-aided composition tool.
The combination of harmonic, rhythmic and melodic composition in a 
single framework makes \systemname{} unique in the growing area of 
algorithmic composition.

With near real-time composition, 
\systemname{} reaches the point where it can not only be used
as a component in an interactive composition tool but also has the potential for live performances and
concerts or automatically generated background music in a variety of applications.
With the use of a fully declarative language and an ``off-the-shelf''
reasoning engine, \systemname{} provides the human composer a tool which is 
significantly simpler, more compact and more versatile than other existing systems.

To appear in Theory and Practice of Logic Programming
(TPLP)
\end{abstract}


\section{Introduction}
Music, although it seeks to communicate via emotions, is almost always
governed by complex and rigorous rules which provide a foundation on
which artistic expression can be based.  In the case of musical
composition, in most styles there are rules which describe the
progression of a melody, both at the local level (the choice of the
next note) and at the global level (the overall structure).  Other
rules describe the harmony, which arises from the relationship between
the melodic line and the supporting instruments. A third set of rules 
determines the rhythm, the intervals between notes, of a piece.

These rules were developed to guide and support human composers
working in the style of their choice, but we wish to demonstrate here
that by using knowledge representation techniques, we can create a
computer system that can reason about and apply compositional
rules. Such a system will provide a simple and flexible way of
composing music automatically. Provided that the representation
technology used is sufficiently flexible to allow changes at the level
of the rules themselves, the system will also help the human composer to
understand, explore and extend the rules (s)he is working with.

Since the beginning of recorded time, music composers have used
a number of processes to generate the next note, be it simple scales or
arpeggios, or complex mathematical structures.  The interest in
developing computational systems for composing, harmonising and
accompanying music is not new either.  Researchers have used a variety of mechanisms
in searching for a viable system, including encodings of the
stochastic and symbolic music of \citeN{Xenakis}, and attempts to find
simpler schemes to the major work in artificial intelligence on the
harmonisation of Bach chorales by \citeN{Ebcioglu}.

This paper describes \systemname, an automatic composition 
system capable of simple melodies with accompaniment, in particular 
for species one counterpoint as practised in the early Renaissance.  The insights
gained from this system can be and are being extended to other musical
styles, by adding or changing the rules.  What has impressed us has
been particularly the ease in which musical experience could be
converted into code which is succinct and easy to verify.

\systemname{} uses Answer Set Programming (ASP) \cite{gellif88}, a logic programming paradigm, 
to represent the music knowledge and rules of the system. A detailed description is provided in
Section~\ref{asp}, after a short description of 
the musical aspects of the project (Section~\ref{music}).  

The initial system, \systemnamev{1.0}, was first presented in \cite{bbdf}.
In this paper, we present \systemnamev{1.5}. The simplicity of the basic encoding is presented 
in Section~\ref{anton}.  As we will demonstrate, the new version excluding rhythm is
significantly faster allowing \systemname{} to be used in real-time application
rather than just as an interactive tool.

The initial system \cite{bbdf}, has only an extremely
simple concept of rhythm, all notes having the same length.  In
Section~\ref{rhythm} we show how this musical restriction has been relaxed,
allowing interesting and ``correct'' rhythmic patterns.  The paper is
completed with a discussion of the performance, both musical and
computational in Section~\ref{results}, the use of ASP in Section~\ref{asp-kr} and
future work in Section~\ref{future}.

Our overall aim is multi-faceted; on the one hand we want develop musicological ideas, create music, and test musical thoughts and on the other can use the system to test the quality and
utility of ASP solvers in real-world applications\footnote{This paper is not intended to be a benchmarking
paper for the various solvers and their numerous option. In this paper we simply wish to demonstrate that ASP 
is an appropriate paradigm for music composition and that composition can be done in near real-time.}, 
We also note that \systemname{}
is usable as part of a student marking system, checking that harmonisations fit the rules.
Alternatively, the system can be used as a diagnostic and assisted composition tool, where a
part of the piece is given to the program to be correctly completed.

\section{Music Theory}
\label{music}

Music is a world-wide phenomenon across all cultures.  The details of
what constitutes music may vary from nation to nation, but it is clear
that music is an important component of being human.

In the work of this paper we are concentrating on western traditional
tonal musics, but the underlying concepts can be translated to other traditions.
The particular area of interest here is composition; that is creating
new musical pieces.

Creating melodies, that is sequences of pitched sounds, is not as easy
as it sounds.  We have cultural preferences for certain sequences of
notes and preferences dictated by the biology of how we hear.  This
may be viewed as an artistic (and hence not scientific) issue, but
most of us would be quick to challenge the musicality of a composition
created purely by random whim.  Students are taught rules of thumb to
ensure that their works do not run counter to cultural norms and also
fit the algorithmically definable rules of pleasing harmony when
sounds are played together.

``Western tonal" simply refers to what most people in the West think
of as ``classical music", the congenial Bach through Brahms, music
which feels comfortable to the modern western ear because of its
adherence to familiar rules.  Students of composition in
conservatoires are taught to write this sort of music as basic
training.  They learn to write melodies and to harmonise given
melodies in a number of sub-versions.  If we concentrate on early
music then the scheme often called informally ``Palestrina Rules'' is an obvious
example to used for such a task.  Similarly, harmonising Bach
chorales is a common student exercise, and has been the subject of
many computational investigations using a variety of methods.

For the start of this work we have opted to work with Renaissance
Counterpoint. This style was used by composers like Josquin, Dufay or
Palestrina and is very distinct from the Baroque Counterpoint used by
composers like Bach, Haendel.

We have used the teaching at one conservatoire in K\"oln to provide
the basic rules, which were then refined in line with the general
style taught.  The point about generating melodies is that the
``tune'' must be capable of being accompanied by one or more other
lines of notes, to create a harmonious whole.  The requirement for the
tune to be capable of harmonisation is a constraint that turns a
simple sequence (a {\em monody}) to a {\em melody}.

In this particular style of music complete pieces are not usually
created in one go.  Composers create a number of sections of melody,
harmonising them as needed, and possibly in different ways, and then
structuring the piece around these basic sections.  Composing between
4 bars and 16 bars is not only a computationally convenient task, it is actually
what the human composer would do, creating components from which the whole is
constructed.  So although the system described here may be 
limited in its melodic scope, it has the potential to become a useful
tool across a range of sub-styles. 

\subsection{Automatic Composition}

A common problem in musical composition can be summarised in the
question ``where is the next note coming from?''.  For many composers
over the years the answer has been to use some process to generate
notes.  It is clear that in many pieces from the Baroque period that
simple note sequences are being elaborated in a fashion we would now
call algorithmic.  For this reason we can say that algorithmic
composition is a subject that has been around for a very long time.
It is usual to credit Mozart's Musikalisches W\"urfelspiel (Musical
Dice Game) \cite{Dicegame} as the oldest classical algorithmic
composition, although there is some doubt if the game form is really
his.  In essence the creator provides a selection of short sections,
which are then assembled according to a few rules and the roll of a
set of dice to form a Minuet\footnote{A dance form in triple time,
  {\em i.e.\/} with 3 beats in each measure}.  Two dice are used to
choose the 16 minuet measures from a set of 176, and another die
selects the 16 trio measures\footnote{A Trio is a short contrasting
  section played before the minuet is repeated}, this time from 96
possibilities.  This gives a total number of $1.3 \times 10^{29}$ possible
pieces.  This system however, while using some rules, relies on the
coherence of the individual measures.  It remains a fun activity, and
recently web pages have appeared that allow users to create their own
original(ish) ``Mozart'' compositions\footnote{For example
\url{http://www.softsynth.com/jsyn/examples/dicegame/},\\
\url{http://jmusic.ci.qut.edu.au/jmtutorial/MozartDiceGame.html},\\
\url{http://www.amaranthpublishing.com/MozartDiceGame.htm},\\
\url{http://imagine.xs4all.nl/bram/mozart},\\
\url{http://magyar-irodalom.elte.hu/kirnberger/_html/mozart_e.html}}.

In the music of the second Viennese school (``12-tone'', serial music)
there is a process of action, rotating, inverting and use of
retrograde, but usually this is performed by hand.

More recent algorithmic composition systems have concentrated on the
generation of monody\footnote{A monody is a single solo line, in
  opposition to homophony and polyphony}, either from a mathematical
sequence, chaotic processes, or Markov chains, trained by
consideration of acceptable other works.  Frequently the systems rely
on a human to select which monodies should be admitted, based on
judgement rather than rules.  Great works have been created this way,
in the hands of great talents.  Probably the best known of the Markov
chain approach is \citeN{Cope}'s significant corpus of Mozart pastiche.

In another variation on this approach, the accompanist, either knowing
the chord structure and style in advance, or using machine-listening
techniques, infers a style of accompaniment.  The former of these
approaches can be found in commercial products, and the latter has
been used by some jazz performers to great effect, for example by
George E. Lewis~\citeyear{Lewis}.

A more recent trend is to cast the problem as one of constraint
satisfaction.  For example PWConstraints is an extension for IRCAM's
Patchwork, a Common-Lisp-based graphical programming system for
composition.  It uses a custom constraint solver employing
backtracking over finite integer domains.  OMSituation and OMClouds
are similar and are more recently developed for Patchwork's successor
OpenMusic.  A detailed evaluation of them can be found in
\cite{AndersPhD}, where the author gives an example of a 1st-species
counterpoint (two voices, note against note) after \cite{Fux}
developed with Strasheela, a constraint system for music built on the
multi-paradigm language Oz.  Our musical rules implement the
melody and counterpoint rules described by \cite{Thakar}, which we
find give better musical results.

One can distinguish between \emph{improvisation} systems and 
\emph{composition} systems.  In the former the note selection progresses
through time, without detailed knowledge of what is to come.  In
practice this is informed either by knowing the chord progression or
similar musical structures \cite{Brothwell}, or using some machine
listening.  In this paper we are concerned with \emph{composition}, so
the process takes place out of time, and we can make decisions in any
order. 

It should also be noted that these algorithmic systems compose pieces of
 music of this style in either a melodic or a harmonic fashion, and are
frequently associated with computer-based synthesis. The system we will 
propose later in this paper is unique as it deals with both simultaneously.

\subsubsection{Melodic Composition}

In melodic generation a common approach is the use of some kind of
probabilistic finite state automaton or an equivalent scheme, which is
either designed by hand (some based on chaotic oscillators or some
other stream of numbers) or built via some kind of learning process.
Various Markov models are commonly used, but there have been
applications of n-grams, genetic algorithms and neural nets.  What
these methods have in common is that there is no guarantee that
melodic fragments generated have acceptable harmonisations.  Our
approach, described below is fundamentally different in this respect,
as our rules cover both aspects simultaneously.

In contrast to earlier methods, which rely on learning, and which are
capable of giving only local temporal structure, a common criticism of
algorithmic melody \cite{Leach}, we do not rely on learning and hence
we can aspire to a more global, whole melody, approach.
Furthermore, learning is designed to work in one direction at the time which makes it hard
to use in a partially automated fashion.
Our systems is not bound to these limitation, so operations
like ``fill
in the 4 notes between these sections'' are not a problem for us.

Researchers are also trying to move beyond experiments with random note
generation because the results are too lacking in structure.  
Predictably, the alternative of
removing the non-determinism at the design stage (or replacing with a
probabilistic choice) runs the risk of `sounding predictable'! There
have been examples of good or acceptable melodies created like this,
but the restriction inherent in the process means it probably works
best in the hands of geniuses.

Our experience with this work made us realise how many acceptable
melodies can be created with only a few rules, and as we add rules,
how much better the musical results are. 

\subsubsection{Harmonic Composition}

A common usage of algorithmic composition is to add harmonic lines to
a melody; that is notes played at the same time as the melody that are
in general consonant and pleasing.  This is exemplified in the
harmonisation of 4-part chorales, and has been the subject of a number
of essays in rule-based or Markov-chain systems.  Perhaps a pinnacle
of this work is \citeN{Ebcioglu} who used early expert system
technology to harmonise in the style of Bach, and was very successful.
Subsequently there have been many other systems, with a range of
technologies.
A review of these is included in \cite{Rohrmeier}.

Clearly harmonisation is a good match to constraint programming based
systems, there being accepted rules\footnote{For example see:
  {\small \url{http://www.wikihow.com/Harmonise-a-Chorale-in-the-Style-of-Bach}}}.
It also has a history from musical education.

But these systems all start with a melody for which at least one valid
harmonisation exists, and the program attempts to find one, which is
clearly soluble.  This differs significantly from our system, as we
generate the melody and harmonisation together, the requirement for
harmonisation affecting the melody. 

\section{Answer Set Programming}
\label{asp}

\textit{Answer Set Programming} \cite{baralbook,gellif88} is a
declarative programming paradigm in which  a logic program is
used to describe the requirements that must be fulfilled by the
solutions of a certain problem.  The answer sets of the program,
usually defined through (a variant/extension of) the stable model
semantics \cite{gellif88}, are interpreted to the solutions of the
problem. This technique has been successfully applied in domains such
as planning \cite{eiter2002,lif2000}, configuration and verification
\cite{soininen99}, super-optimisation \cite{bcdvf06b}, diagnosis
\cite{eiter99}, game theory \cite{dvver99a} multi-agent
systems \cite{baral2000,bucca2002,dvver04,buccafurri+caminiti:2005,cdvp06c},
reasoning about biological networks \cite{grscse06b}, 
voting theory \cite{konczak06b}, policy mechanisms~\cite{milsch06a},
generation of phylogenetic trees \cite{ErdemLR06}, evolution of
language \cite{ErdemLNR03} and game character descriptions
\cite{PadovaniP04}.

There is a large body of literature on ASP:
for in-depth coverage see \cite{baralbook}, but for the sake
of making this paper self-contained we will cover the essentials as
they pertain to our usage here.

\paragraph{Basic Concepts:}
The answer set semantics is a model based semantics for normal logic programs.
Following the notation of \cite{baralbook}, we refer to the language of normal logic programs
under the answer set semantics as \AnsPrologs.

The smallest building block of an \AnsPrologs{} program is an atom or
predicate, e.g. $\asp{r(X,Y)}$ denotes $X r Y$.
e.g \asp{owns(X,Y)} stating that $X$ owns $Y$.  
$X$ and $Y$ are variables which can be grounded with constants, e.g. owns(alice,key).  Each
ground atom can be assigned the truth value {\em true} or {\em false}.

In this paper we will only consider programs with one type of
negation, namely negation-as-failure denoted ${\mathbf{not}}$. This
type of negation states that something should be assumed false when it
cannot be proven to be true. A literal is an atom $a$ or its negation
$\NAF a$, with $\NAF \NAF a = a$.  
We extend the notation to sets: $\NAF S$ is the set $\{\NAF
l \mid l \in S\}$ with $S$ a set of literals.

An \AnsProlog program consist of a finite set of statements, called
rules.  Each rule $r: a\gets B.$ or $ \bot \gets B.$ is made of two
parts namely the body $B$, denoted $B_r$, which is a set of literals,
and a head atom $a$ or $\bot$, denoted $H_r$. The body can be divided
in two parts: the set of positive atoms, denoted as $B^+_r$, and the
set of negated atoms, denoted $B^-_r$.  A rule should be read as:
``$a$ is {\em supported\/} if all elements of $B$ are true''.  A rule
with empty body is called a {\em fact} and we often only mention the
head. A rule with head $\bot$ is referred to as an {\em (integrity)
constraint}. We often omit the $\bot$ symbol and leave the head
empty. $\bot$ is always assigned the truth value ``false''.  A program
is called positive if it does not contain any negated literals.

The finite set of all constants that appear in the program $P$ is
referred to as the {\em Herbrand Universe}, denoted $\universe{P}$.
Using the Herbrand universe, we can ground the entire program. Each
rule is replaced by its ground instances, which can be obtained by
replacing each variable symbol by an element of $\universe{P}$. The
ground program, denoted $\ground{P}$, is the union of all ground
instances of the rules in $P$.

The set of all atoms grounded over the Herbrand universe of a program
is called the {\em Herbrand Base}, denoted as $\hbase{P}$.  These are
exactly those atoms that will appear in the ground program.

An assignment of truth values to all atoms in the program (or all
elements from the Herbrand base) is
called an {\em interpretation}. Often only those literals that are
considered true are mentioned, as all the others are false by
definition (negation as failure).

Given a ground rule $r$, we say a $r$ is {\em applicable} w.r.t. an
interpretation $I \subseteq \hbase{P}$ if all the body elements are
true ($B^+_r \subseteq I$ and $B^-_r \cap I = \emptyset$). The rule
is {\em applied} w.r.t. $I$ when it is applicable and $H_r \in I$.  A
ground rule is {\em satisfied} w.r.t. an interpretation $I$ if it is
either not applicable or applied w.r.t. $I$.  An atom is {\em
supported} w.r.t. $I$ if there is an applied rule with this atom in
the head.
Obviously, we want to make sure that interpretations satisfy every
rule in the program. So, an interpretation $I$ is a {\em model} for a
program $P$ if and only if all rules in $\ground{P}$ are satisfied.

To find actual solutions, models alone are not sufficient:
we need to make sure that only those literals that are supported
are considered true. This results in the so-called minimal model
semantics. A model $M$ for a program $P$ is {\em minimal} if no other
model $N$ exists such that $N \subset M$.  Programs can have any
number of minimal models, while programs without constraints will
always admit at least one. Positive programs will have at most one minimal model,
and exactly one when they do not admit any constraints.

The minimal model of a positive program without constraints can be
found using a fixpoint, called the {\em deductive closure}, which can
be computed in polynomial time. We start with the empty set and find
all atoms that are supported. With this new set we continue to find
supported atoms. When the set reaches a fixpoint, we have found the
deductive closure or minimal model of the program. When the program  
contains constraints, we can follow the same principle
but the process fails when a unsatisfied constraint is found.

\begin{definition}\label{def:tp}
Let $P$ be a positive \AnsProlog program and let $I$ be an
interpretation. We define the {\em immediate consequence operator}
$\TPR$ as:
\[
\TP{I} = \{a \in \hbase{P} \mid \Exists{r \in P}{B_r \subseteq I}\}
\]
\end{definition}

$\TPR$ is monotonic so it has a least fixpoint, denoted $\TPL$, which corresponds to the deductive closure.

\begin{definition}
Let $P$ be a positive \AnsProlog program. The \textit{deductive closure} of $P$,
is the least fixpoint $\TPL$ of $\TPR$.
\end{definition}

The minimal semantics is sufficient for positive programs, but fails in the presence of 
negation-as-failure.  A simple example of such a program is: $\{\asp{a \gets
\NAF a.};\asp{b \gets a}\}$. This program has one minimal model
$\{a,b\}$, while the truth of $a$ depends on $a$ being false.
To obtain intuitive solutions, we need to verify that our assumptions are
indeed correct. This is done by reducing the program to a simpler
program containing no negation-as-failure. Given an
interpretation, all rules that contain $\NAF l$ that are considered
false are removed while the remaining rules only retain their
body atoms. This reduction is often referred to as the Gelfond-Lifschitz
transformation \cite{gellif88,gellif91}.  When this program gives the
same supported literals as the ones with which we began, we have found
an answer set.

\begin{definition}
Let $P$ be a ground \AnsProlog program. The \textit{Gelfond-Lifschitz
transformation} of $P$ w.r.t a set of ground atoms $S$,
is the program $P^S$ containing the rules $H_r \gets B^+_r$ such that $H_r \gets B_r^+, \NAF B_r^- \in P$ with $B_r^- \cap S = \emptyset$.
\end{definition}

\begin{definition}
Let $P$ be a \AnsProlog. A set of ground 
atoms $S \subseteq \hbase{P}$ is an \textit{answer set} of $P$ iff $S$
is the minimal model of $\ground{P^S}$.
\end{definition}

The non-deterministic nature of negation-as-failure gives rise to several
answer sets, which are all acceptable solutions to the problem that
has been modelled. It is in this non-determinism that the strength of
answer set programming lies.

\paragraph{Extensions}
The basic language, single head atom and negation-as-failure only 
appearing in the body, already enables the representation of many problems.
However, for some applications the programmer is forced to write code in a more
round-about non-intuitive way. To overcome this, several extensions were introduced.

For this paper we use one of these extensions:  choice rules \cite{NiemelaSS99}.
A lot of problems require choices to made between a set of atoms. 
Although this can be modelled in the basic formalism
it tends choice rules are offer more clarity of expression and are more convenient.
Choice rules, written $L \{l_1, \ldots, l_n\} M$, are a
convenient construct to state that at least $L$ and at most $M$
from the set $\{l_1 , \ldots, l_n\}$ must be true in order to satisfy
the construct.  $L$ defaults to 0 when omitted while $M$ defaults to $n$. Choice
rules are often used in conjunction with a grounding predicate: $L
\{A(X):B(X)\} M$ represents the choice of a number of atoms $A(X)$
where is grounded with all values of $X$ for which $B(X)$ is true.

\paragraph{Implementations:}
Algorithms and implementations for obtaining answer sets of logic
programs are referred to as {\em answer set solvers}.  The most
popular and widely used solvers are \textsc{smodels} \cite{nisi97} and \textsc{dlv} \cite{eilemapfsc98} 
and more recently \textsc{clasp} \cite{gekanesc07a}.

Alternatives are \textsc{cmodels} \cite{sat-asp} and \textsc{sup}
\cite{sup}, solvers based on translating the program to a SAT problem,
and \textsc{smodels-{ie}} \cite{smodels-ie}, the cache-efficient version
of \textsc{smodels}.  Furthermore, there is the distributed solver
\textsc{platypus} \cite{grjamescthti05a}.

To solve a problem it first needs to be grounded. Currently three
grounders are used: the grounder integrated with \textsc{dlv}, 
\textsc{lparse}, the grounder that was developed together with
\textsc{smodels} but is used by most solvers, and the most recent one
\textsc{GrinGo} \cite{gescth07a} which works together with \textsc{clasp}
and other solvers that take \textsc{lparse} output. 
During the grounding phase, not only are the variables substituted for
constants, but also useless rules are
eliminated. Furthermore, grounders try to simplify the program as
much as possible. The second phase is solving which takes a grounded
program or its internal representation as input and generates the set of its answer sets.

Current answer set solvers can be divided in three groups depending on
the style of algorithm they use or the mapping the use:
depth first search (DPLL) 
clause learning 
and a mapping to SAT.
All use a variety of heuristics to improve the
performance of the basic algorithm.

\section{Anton}\label{anton}

\begin{figure}[tbp]
\hspace{-1cm}
\scalebox{0.7}{
\begin{ttfamily}
\begin{tabular}{l}
\%\% The number of parts is given by the style
\#domain part(P).\\
\%\% Each part picks at most one note per time step\\
\%\% the length of the sequence is provided at run-time\\
time(1..t).\\
\#domain time(T).\\
\mbox{}\\
\%\% Each part can only play one note at a given time\\
\ \ :- 2 \{ chosenNote(P,T,NN) : note(NN), rest(P,T) \}.\\
\mbox{}\\
\%\% At every time step the note may change\\
\%\% It changes by stepping (moving one note in the scale)\\
\%\% or leaping (moving more than one note)\\
\%\% These can either be upwards or downwards\\
1 \{ changes(P,T), repeated(P,T), toRest(P,T), fromRest(P,T), \\
\ \  incorrectProgression(P,T) \} 1 :- T != t.\\
1 \{ stepAt(P,T), leapAt(P,T) \} 1 :- changes(P,T), T != t.\\
1 \{ downAt(P,T),   upAt(P,T) \} 1 :- changes(P,T), T != t.\\
\mbox{}\\
stepDown(P,T) :- stepAt(P,T), downAt(P,T).\\
stepUp(P,T)   :- stepAt(P,T),   upAt(P,T).\\
\mbox{}\\
\#const err\_ip="Incorrect progression".\\
reason(err\_ip).\\
error(P,T,err\_ip) :- incorrectProgression(P,T).\\
\mbox{}\\
\%\% If we step, we must pick an amount to step by\\
1 \{ stepBy(P,T,SS) : stepSize(SS) : SS < 0 \} 1 :- stepDown(P,T).\\
1 \{ stepBy(P,T,SS) : stepSize(SS) : SS > 0 \} 1 :- stepUp(P,T).\\
\mbox{}\\
\%\% Make it so\\
chosenNote(P,T + 1,N + S) :- chosenNote(P,T,N), stepAt(P,T), stepBy(P,T,S), note(N + S).\\
chosenNote(P,T + 1,N + L) :- chosenNote(P,T,N), leapAt(P,T), leapBy(P,T,L), note(N + L).\\
chosenNote(P,T + 1,N) :- chosenNote(P,T,N), repeated(P,T).\\
\end{tabular}
\end{ttfamily}
}

\caption{A code fragment from \filename{progression.lp}}\label{code:progression}
\end{figure}

\subsection{System Description}
\systemname{ } is an algorithmic composition system that uses ASP techniques to represent and reason
about compositional rules.  \AnsProlog is used to
write a description of the rules that govern the melodic and harmonic
properties of a correct piece of music; in this way the program describes
a model of musical composition that can be used to assist the
composer by suggesting, completing and verifying short pieces.
The composition rules are modelled 
so that the \AnsProlog program defines the requirements for a piece to
be musically valid, and thus every answer set corresponds to a
different valid piece.  To generate a new piece the composition system
simply has to generate an (arbitrary) answer set.

In this section we will discuss the basic system of \systemnamev{1.5}. 
The handling of rhythm will be discussed in Section~\ref{rhythm}
page \pageref{rhythm}.

The \AnsProlog program of the basic system is created from various
files: \filename{notes.lp}, \filename{modes.lp},
\filename{progression.lp}, \filename{melody.lp}, \filename{harmony.lp}
and \filename{chord.lp}. The first contains the general background rules on notes
and intervals while the second describes the various modes/keys the
system can use and their consequences for note selection and position.
The current system is able to work with major, minor, Dorian, Lydian
and Phrygian modes.  Rules for the
progression of all parts, either melodic and harmonic, are handled in
\filename{progression.lp}.  This part of the program is responsible for
selecting the next note in each of the parts on the basis of the
previous note.  The rules for the melodic parts and for composing with
multiple parts are encoded in \filename{melodic.lp} and \filename{harmonic.lp}
respectively.  \filename{chord.lp} provides the description of chords and
chordal progression and the effects of note choices.

We will discuss progression, melody and harmony in more detail.
The whole system is licensed under the GPL and publicly
available\footnote{The source code is available from \url{http://www.cs.bath.ac.uk/~mjb/anton/}}.

Figure~\ref{code:progression} presents a selection of rules dealing
with the progression of notes.  The model is defined over a number of time
steps, given by the variable \texttt{T}.  The key proposition is
\texttt{chosenNote(P,T,N)} which represents the concept ``At time
\texttt{T}, part \texttt{P} plays note \texttt{N}''.  To encode the
options for melodic progress (``the tune either steps up or down one
note in the key, leaps more than one note, repeats or rests''), choice
rules are used.  For diagnostic and debugging purposes,  
compositional errors are not immediately encoded as constraints, but instead
use error rules like \texttt{error(P,T,err\_ip) :-
  incorrectProgression(P,T)}.  By using a constraint to obtain answer
sets with error-atoms or excluding them entirely, we can alter the
functionality of our system from diagnosis to composition without changing the code.  We will later
return to various applications of our system.

\begin{figure}[tbp]
\hspace{-1cm}
\scalebox{0.60}{
\begin{ttfamily}
\begin{tabular}{l}
\%\% Melodic parts are not allowed to repeat notes\\
\#const err\_nrmp="No repeated notes in melodic parts".\\
reason(err\_nrmp).\\
error(MP,T,err\_nrmp) :- repeated(MP,T).\\
\mbox{}\\
\%\% A leap of an octave is only allowed from the fundamental.\\
\#const err\_olnf="Leap of an octave from a note other than the fundamental".\\
reason(err\_olnf).\\
error(MP,T,err\_olnf) :- leapBy(MP,T,12), not chosenChromatic(MP,T,1).\\
error(MP,T,err\_olnf) :- leapBy(MP,T,-12), not chosenChromatic(MP,T,1).\\
\mbox{}\\
\%\% Impulse\\
\%\% Stepwise linear progression creates impulse\\
\%\% Leaps create impulse - using the notes in between resolves this\\
downwardImpulse(MP,T+1) :- leapDown(MP,T), time(T+1).\\
downwardImpulse(MP,T+3) :- stepDown(MP,T+2), stepDown(MP,T+1), stepDown(MP,T), time(T+3).\\
upwardImpulse(MP,T+1) :- leapUp(MP,T), time(T+1).\\
upwardImpulse(MP,T+3) :- stepUp(MP,T+2), stepUp(MP,T+1), stepUp(MP,T), time(T+3).\\
\mbox{}\\
\%\% No repetition of two or more notes\\
\#const err\_rn="Repeated notes".\\
reason(err\_rn).\\
error(MP,T1,err\_rn) :- chosenNote(MP,T1,N), stepBy(MP,T1,S1),\\
\mbox{}\hspace{2cm}    chosenNote(MP,T2,N), stepBy(MP,T2,S1),\\
\mbox{}\hspace{2cm}    T1 + 1 < T2, T2 < T1 + 2 + RW.\\
error(MP,T1,err\_rn) :- chosenNote(MP,T1,N), leapBy(MP,T1,L1),\\
\mbox{}\hspace{2cm}    chosenNote(MP,T2,N), leapBy(MP,T2,L1),\\
\mbox{}\hspace{2cm}    T1 + 1 < T2, T2 < T1 + 2 + RW.\\
\#const err\_dc="Dissonant contour".\\
reason(err\_dc).\\
error(MP,t,err\_dc) :- lowestNote(MP,N1), highestNote(MP,N2),\\
\mbox{}\hspace{2cm}   chromatic(N1,C1),  chromatic(N2,C2),\\
\mbox{}\hspace{2cm}   not consonant(C1,C2), N1 < N2.\\
\end{tabular}
\end{ttfamily}
}

\caption{A code fragment from \filename{melody.lp}}\label{code:melody}
\end{figure}

To encode the melodic limits on the pattern of notes
and the harmonic limits on which combinations of notes may be played
at once, error-rules like the one in \filename{progression.lp} are included. 
Figure~\ref{code:melody} shows how we encoded rules that forbid repetition of
notes in the melodic parts, octave leaps except for special circumstances, 
impulses, repetition of more than two notes and certain lengths of intervals.
While some of these rules might be valid in other types of music,
Renaissance counterpoint explicitly forbids them.

Interaction between parts is governed by the rules of harmony. 
Figure~\ref{code:harmony} shows how we encoded the musical rules that 
specify that you cannot have dissonant intervals between parts,
that limit the distance between parts and that state that parts cannot cross-over.

\begin{figure}[tbp]
\hspace{-1cm}
\scalebox{0.7}{
\begin{ttfamily}
\begin{tabular}{l}
\#const err\_dibp="Dissonant interval between parts".\\
reason(err\_dibp).\\
error(P1,T,err\_dibp) :- chosenChromatic(P1,T,C1), chosenChromatic(P2,T,C2),\\
\mbox{}\hspace{2cm}  P1 < P2, chromaticInterval(C1,C2,D),\\
\mbox{}\hspace{2cm}  not validInterval(D).\\
\mbox{}\\
\%\% The maximum distance between parts is an octave plus 4 semitones (i.e. 16 semitones).\\
\#const err\_mdbp="Over maximum distance between parts".\\
reason(err\_mdbp).\\
error(P,T,err\_mdbp) :- chosenNote(P,T,N1), chosenNote(P+1,T,N2),\\
\mbox{}\hspace{2cm}  N1 > N2 + 16, part(P+1).\\
\mbox{}\\
\%\% Parts cannot cross over.\\
\#const err\_pcc="Parts can not cross".\\
reason(err\_pcc).\\
error(P,T,err\_pcc) :- chosenNote(P,T,N1), chosenNote(P+1,T,N2),\\
\mbox{}\hspace{2cm} N1 < N2, part(P+1).\\
\end{tabular}
\end{ttfamily}
}

\caption{A code fragment from \filename{harmony.lp}}\label{code:harmony}
\end{figure}

While the fragments shown in Figures
\ref{code:progression}-\ref{code:harmony} are a subset of the knowledge base,
 they demonstrate that the rules are very simple and intuitive
(with the necessary musical background). The modelling of this style
of music, excluding rhythm, contains less than 400 ungrounded logic
rules.  

\subsection{Features}
In the previous section we discussed the basic components of the \systemname{} system.
However, in order to have a complete system, we are still missing one
component: the specification of parts.  Currently, the system comes
with descriptions for solos, duets, trios and quartets but the basic
system is written with no fixed number of parts in mind.
Figure~\ref{code:quartet} shows the description for a quartet. 

\begin{figure}[tbp]
\hspace{-1cm}
\scalebox{0.7}{
\begin{ttfamily}
\begin{tabular}{l}
\%\% This is a quartet\\
style(quartet).\\
\mbox{}\\
\%\% There are four parts\\
part(1..4).\\
\mbox{}\\
\%\% The top part plays the melody\\
melodicPart(1).\\
\mbox{}\\
\%\% For chords we need to know the lowest part\\
lowestPart(4).\\
\mbox{}\\
\%\% We need a range of up to 2 octaves (24 steps) for each part,\\
\%\% thus need 24 notes above and below the lowest / highest start\\
\#const quartetBottomNote=1.\\
\#const quartetTopNote=68.\\
note(quartetBottomNote..quartetTopNote).\\
bottomNote(quartetBottomNote).\\
topNote(quartetTopNote).\\
\mbox{}\\
\%\% Starting positions are 1 - 5 - 1 - 5\\
\#const err\_isn="Incorrect starting note".\\
reason(err\_isn).
error(1,1,err\_isn) :- not chosenNote(1,1,44).\\
error(2,1,err\_isn) :- not chosenNote(2,1,37).\\
error(3,1,err\_isn) :- not chosenNote(3,1,32).\\
error(4,1,err\_isn) :- not chosenNote(4,1,25).\\
\mbox{}\\
\%\% No rests\\
\#const err\_nrfw="No rest for the wicked".\\
reason(err\_nrfw).\\
error(P,T,err\_nrfw) :- rest(P,T).\\
\mbox{}\\
\%\% With three or more parts allow intervals of a major fourth\\
\%\% (5 semitones) between parts\\
validInterval(5).\\
\end{tabular}
\end{ttfamily}
}

\caption{The quartet specification}\label{code:quartet}
\end{figure}

Depending on how the system is used, for composition or diagnosis,
one will either be interested in those pieces that do not result in
errors at all, or in an answer set that mentions the error messages. 
For the former we simply specify the constraint 
{ \texttt{:- error(P,T,R).}}, effectively making any error rule
into a constraint. For the latter we include the
rules: { \texttt{errorFound :- error(P,T,R).}} and 
{ \texttt{:- not errorFound.}} requiring that an error is found
({\em i.e.\/} returning no answers if the diagnosed piece is error free).
These simple rules are encoding in \filename{composing.lp} and \filename{diagnosis.lp} so that they can be included when our scripts assemble to program.

By adding constraints on which notes can be included, it is possible
to specify a part or all of a melody, harmony or complete piece.  This
allows \systemname{ } to be used for a number of other tasks beyond
automatic composition.  By fixing the melody it is possible to use it
as an automatic harmonisation tool.  By fixing part of a piece, it can
be used as computer aided composition tool.  By fixing a complete
piece, it is possible to check its conformity with the rules, for
marking student compositions or harmonisations.  Alternatively we
could request the system to complete part of a piece.  

The complete system consists of three major phases; building the
program, running the \AnsProlog program and interpreting the results.  As a
simple example suppose we wish to create a 4 bar piece in E major. One
would write:

{\small
\begin{verbatim}
programBuilder.pl --task=compose --mode=major --time=16 > program
\end{verbatim}
}

\noindent which builds the \AnsProlog program, giving the length and mode.  Then

{\small
\begin{verbatim}
gringo < program | ./shuffle.pl 6298 | clasp 1 > tunes
\end{verbatim}
}

\noindent runs the solving phase and generates a representation of the piece.
We provide a number of output formats, one of which is a \prog{Csound} \cite{csound}
program with a suitable selection of sound templates.

{\small
\begin{verbatim}
$ parse.pl --fundamental=e --output=csound < tunes > tunes.csd
\end{verbatim}
}

\noindent generates the \prog{Csound} input from the generic format, and then

\begin{figure*}[t]
\begin{center}
\includegraphics[width=0.8\textwidth]{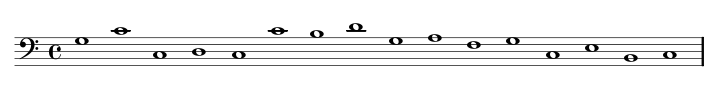}

\begin{verbatim}
         | 55 60 48 50 48 60 59 62 55 57 53 55 48 52 47 48 
         |  G  C  C  D  C  C  B  D  G  A  F  G  C  E  B  C 
         |   +5 -12 +2 -2 +12 -1 +3 -7 +2 -4 +2 -7 +4 -5 +1 
\end{verbatim}
\end{center}
\caption{The score and human readable format for the tunes composition.}\label{score}
\end{figure*}

{\small
\begin{verbatim}
$ csound tunes.csd -o dac
\end{verbatim}
}

\noindent plays the melody.
We provide in addition to \prog{Csound}, output in human readable format, \AnsProlog facts or the
Lilypond score language.
Figure~\ref{score} shows the score of the tunes piece composed above.

Alternatively we could request the system to complete part of a piece.
In order to do so, we provide the system with a set of \AnsProlog facts
expressing the mode, the notes which are already fixed, the number
of notes in your piece, the configuration and the number of parts.

Figure~\ref{partial} contains an example of such file.

\begin{figure}[b]
\begin{center} \scalebox{0.9}{
\begin{ttfamily}
\begin{tabular}{|l}
keyMode(lydian).\\
chosenNote(1,1,25).\\
chosenNote(1,2,24).\\
chosenNote(1,8,19).\\
chosenNote(1,9,20).\\
chosenNote(1,10,24).\\
chosenNote(1,14,29).\\
chosenNote(1,15,27).\\
chosenNote(1,16,25).\\
\#const t=16.\\
style(solo).\\
part(1).
\end{tabular}
\end{ttfamily}}
\end{center}
\vspace{-10pt}
\caption{\filename{musing.lp}: An example of a partial piece}\label{partial}
\end{figure}

We then use the system as before with the exception of adding
{ \texttt{--piece=musing.lp}} when we run { \texttt{programBuilder.pl}}. The
system can then return all possible valid composition that satisfy
the criteria set out in the partial piece.

\begin{figure}[bt]
\begin{center} \scalebox{0.9}{
\begin{ttfamily}
\begin{tabular}{|l}
mode(major).\\
chosenNote(1,1,25).\\
chosenNote(1,2,24).\\
chosenNote(1,3,29).\\
chosenNote(1,4,27).\\
chosenNote(1,5,32).\\
chosenNote(1,6,30).\\
chosenNote(1,7,34).\\
chosenNote(1,8,32).\\
chosenNote(1,9,34).\\
chosenNote(1,10,39).\\
chosenNote(1,11,37).\\
chosenNote(1,12,34).\\
chosenNote(1,13,36).\\
chosenNote(1,14,37).\\
\#const t=14.\\
style(solo).\\
part(1).
\end{tabular}
\end{ttfamily}}
\end{center}
\vspace{-10pt}
\caption{\filename{problems.lp}: An example of a broken piece}\label{diagnosis}
\end{figure}

\systemname{} is not limited to the composition of music; it can also be used verifying whether
a (partial) piece of music adheres to the rules of compositions. Take for example the composition in Figure~\ref{diagnosis}.
Using the command-line options 
{ \texttt{--task=diagnosis --piece=problems.lp}} when we run { \texttt{programBuilder.pl}}, the system will
return the error codes for all the rules that the composition breaks.
In case of \filename{problems.lp}, we obtain:
{\small
\begin{verbatim}
error(1,2,"Repeated pattern") error(1,4,"Repeated pattern") 
error(1,2,"Split melody") 
\end{verbatim}
}

The first two errors indicate that we have repeated patterns:
e.g. from the second note we leap up by a perfect fourth and step down
by a major second.  This pattern is then repeated starting at the
fourth and sixth notes. The repeated progression pattern also triggers the split melody error. Split melodies occur when the even/odd notes form separate melodies.

The \AnsProlog programs used in \systemnamev{1.5 }, contains just 509 lines
(not including comments and empty lines) and encodes 44 melodic,
harmonic and rhythmic rules.  Once instantiated, the generated programs range from
1,000 atoms and 5,500 rules (a solo piece with 4 notes) to 22,000
atoms and 796,000 rules (a 32 note quartet).
It should be noted that our 1500 lines of code, \AnsProlog rules and
perl scripts combined, contrast with the 8000 lines in Strasheela
\cite{AndersPhD} and 88000 in Bol \cite{Bol}. 

The question of problem complexity is still open.  Clearly there is an
NP algorithm for generating a piece that meets a set of musical
rules and there is no known polynomial algorithm for this task.
Depending on what constitutes a valid musical style, it is likely that
the task can be shown to be NP-Complete. Verifying if a certain composition 
satisfies all the musical rules is linear. However from an applications
standpoint, the exact complexity of the process is of less importance
than the run time performance.

\section{Rhythm}
\label{rhythm}

Our system used to be limited in terms of one of the most essential
musical parameters: rhythm. All music \systemnamev{1.0} generated 
was based on rules for classical polyphony, {\em i.e.\/} the combination of
musically unique, independent melodic voices, but all events had the
same time interval. Within this restriction we are able to generate
first-species counterpoint up to four independent voices and solo
melodies as well as homophonic 4-part chorales where the rules for the
inner voices could be more relaxed from the strict melodic rules that
determine each of the four voices in a polyphonic setting.

Rhythm determines at what point in time a particular note in a
sequence is played, for how long it is played and how much stress this
note will get. The musical experience is partly determined by the
constant inter-change of \emph{impact} and \emph{resolution}. Both
influence the behaviour of musical parameters on micro- and
macro-structural levels and are expressed within the rhythm of the
piece.

An alternative characterisation of rhythm is as the interplay between
the duration of notes.  The relation between notes can then be
expressed in terms of interval relationships \citeN{Allen}.  Thus
temporal interval logic \citeN{Goranko} style reasoning can be used to
describe a constructive model of rhythm.  Our experience has shown
that access to the full range of interval relations is unnecessary and
difficult to encode, thus \systemname{} uses a simpler mechanism, that collects
many of the relationships into one ``overlaps'' literal.  This is
sufficient for the rhythms of our musical genre and it is this simpler
form that we describe here.

Our representation of rhythm is closely related to 
Farey sequence. The Farey sequence of order n can defined as the sequence of reduced fractions in the range [0,1], when
in lowest terms, have denominators less than or equal to n, arranged in order of increasing size.
The $n^{\rm th}$ term in the sequence is the set
of rational numbers with $n$ as denominator. 
An example the sequence of order 8 is
\[
F_8 = {\frac{0}{1}, \frac{1}{8}, \frac{1}{7}, \frac{1}{6}, \frac{1}{5}, \frac{1}{4}, \frac{2}{7}, \frac{1}{3}, \frac{3}{8}, \frac{2}{5}, \frac{3}{7}, \frac{1}{2}, \frac{4}{7}, \frac{3}{5}, \frac{5}{8}, \frac{2}{3}, \frac{5}{7} \frac{3}{4}, \frac{4}{5}, \frac{5}{6}, \frac{6}{7}, \frac{7}{8}, \frac{1}{1}}
\]
The related concept we will use is the filtered Farey sequence, which is a systematic
filtering of this sequence.  In particular we will show how it is
used in the description of rhythm in the next subsection, and indicate
the constructive programming aspects in Section~\ref{FareyProg}.

\subsection{Musical Discussion of Rhythm}

The musical areas of harmony and
classical counterpoint often have rules about the conditions when a
particular constellation of notes may occur on the timeline, sometimes
with inclusion of the events preceding and following. For example, a
\emph{suspended fourth} is a dissonant constellation of two or more
voices created on a strong beat. The voice that suspends the consonant
note, {\em i.e.\/} the fourth note of the scale suspending the third, needs to
be sounding before the dissonance occurs. This sound needs also to be
consonant. After the dissonance on the strong beat the suspending
voice needs then to resolve on a weak time interval within the
measure. That sound again must be a consonance. To help expressing
such rules the interval relations of `start', `end', `during' and
`overlap' are needed. In music, however, the time
intervals themselves are never `neutral'. There are typical
alternations of strong and weak beats in nearly every musical style
that is based on an underlying pulse. Those alternating beat patterns
form the notion of meter. But, the concept of meter in Renaissance
style is a particular one and not the same as the concept of bars,
beats and time signatures that is prevailing since the Baroque era
and which is still in use today.

There are generally four different kinds of musical time in the
Renaissance.  When looking at the subdivision of the brevis, which
translates into today's double whole-note, there are four different
options for the composer of that area, as documented in the famous
treatise `Ars Nova' by Philippe de Vitry. The brevis can be
interpreted either as 3 or 2 semi-breve (today's whole note) and those
further into 3 and 2 subdivisions called minims (today's half note),
so we can subdivide the longa into \\3 x 2, 2 x 2, 3 x 3 or 2 x 3
minims\footnote{The order of subdivision is musically important in
  terms of accentuation and therefore has consequences for the
  treatment of consonance and dissonance. Hence, 3 x 2 is not the same
  as 2 x 3.}. Those different subdivisions were indicated in the vocal
score using different time-signatures\footnote{The four time
  signatures are called \emph{tempus perfectum cum prolationem
    imperfecta} (3 x 2), \emph{tempus imperfectum cum prolationem
    imperfecta} (2 x 2), \emph{tempus perfectum cum prolationem
    perfecta} (3 x 3) and \emph{tempus imperfectum cum prolationem
    perfecta} (2 x 3)}. The 15th century composer Johannes Ockeghem
wrote his \emph{Missa prolationem}, a four-voiced polyphonic
masterpiece, using all four different time-signatures together, a
different one for each voice, while at the same time the voices
imitate their lines containing the same melodic material using canons
and double-canons\footnote{A canon is a special case of a line that
  can be imitated simultaneously by another voice $n$ beats later
  while their combination at the same time satisfies the rules on
  consonance/dissonance. Adding support for more sophisticated interval 
relations in future versions of \systemname will allow us to
 compose canons.}. This complex
polyrhythmic structure establishes a proportion of note durations as
6:4:9:6 between the voices. The effect of superimposing different beat
qualities can be seen in Table \ref{tab:2}. Only the smaller note
durations below the minim are not affected by the proportional
scaling. They remain at the same length for all four voices regardless
of their time-signature. When transcribed nowadays into common
practice notation the voices would need 36 half notes to complete one
large cycle of accent patterns before starting over with a common
`downbeat' \cite{ddlm:kp}. The complete cycle therefore translates
into the filtered Farey Sequence $F_{36}'$, where each of the ratios
in the range of $[0...1]$ denotes the onset time of a half note; see
the representation for one voice, the Bass, in Table
\ref{tab:1}. According to the time-signature (3 x 2) of that
particular voice there are three distinct metrical layers that govern
the note events occurring on such a grid: The top layer gives us the
occurrence of the downbeat, which is the event of most metrical
importance, {\em i.e.\/} no dissonance may occur here unless the voice
causing it prepares the note on the preceding beat. The second layer
denotes the beat level. Although of less metrical weight, no dissonant
interval should sound here unless it is a prepared suspension. The
third layer provides the lightest metrical events. Here dissonances
may occur, for instance in form of a passing note. The rule demanding
preparation of the dissonance is relaxed on this level and on all
further subdivisions underneath.

\begin{table}
\begin{center}

\begin{tabular}{c c c c c c c c c c c c c c c c c c c c}
\toprule
  & \multicolumn{18}{l}{`X' marks a downbeat, `O' marks a 2nd level beat, `o'
    marks a 3rd level beat} \\
\cmidrule(rl){2-20}
S &X &o &O &o &O &o &X &o &O &o &O &o &X &o &O &o &O &o & \\
A &X &o &O &o &X &o &O &o &X &o &O &o &X &o &O &o &X &o & \\
T &X &o &o &O &o &o &O &o &o &X &o &o &O &o &o &O &o &o & \\
B &X &o &o &O &o &o &X &o &o &O &o &o &X &o &o &O &o &o & \\
\midrule
S &X &o &O &o &O &o &X &o &O &o &O &o &X &o &O &o &O &o &X\\
A &O &o &X &o &O &o &X &o &O &o &X &o &O &o &X &o &O &o &X\\
T &X &o &o &O &o &o &O &o &o &X &o &o &O &o &o &O &o &o &X\\
B &X &o &o &O &o &o &X &o &o &O &o &o &X &o &o &O &o &o &X\\
\bottomrule
\end{tabular}
\caption{Polyrhythmic structure of downbeats from Ockeghem's Missa
  prolationem creating a hyper-meter of 36 beats}
\label{tab:2}
\end{center}
\end{table}

\begin{table}[t!]
\scalebox{0.9}{
\begin{minipage}{14cm}
{\footnotesize{
\begin{tabular}{cllllll|llllll|llllll}
\toprule
  & \multicolumn{18}{l}{Tempus perfectum cum prolatione imperfecta - 6 measures x 3 x 2 half notes} \\
\cmidrule(rl){2-19}
I & $\frac{0}{1}$ & & & & & & $\frac{1}{6}$ & & & & & & $\frac{1}{3}$ & & & & & \\
II & & & $\frac{1}{18}$ & & $\frac{1}{9}$ & & & & $\frac{2}{9}$ & &
$\frac{5}{18}$ & & & & $\frac{7}{18}$ & & $\frac{4}{9}$ & \\
III & & $\frac{1}{36}$ & & $\frac{1}{12}$ & & $\frac{5}{36}$ & & $\frac{7}{36}$ & &
$\frac{1}{4}$ & & $\frac{11}{36}$ & & $\frac{13}{36}$ & & $\frac{5}{12}$ & &
$\frac{17}{36}$ \\
\midrule
I & $\frac{1}{2}$ & & & & & & $\frac{2}{3}$ & & & & & & $\frac{5}{6}$ & & & & & \\
II & & & $\frac{5}{9}$ & & $\frac{11}{18}$ & & & & $\frac{13}{18}$ & &
$\frac{7}{9}$ & & & & $\frac{8}{9}$ & & $\frac{17}{18}$ & \\
III & & $\frac{19}{36}$ & & $\frac{7}{12}$ & & $\frac{23}{36}$ & & $\frac{25}{36}$ & &
$\frac{3}{4}$ & & $\frac{29}{36}$ & & $\frac{31}{36}$ & & $\frac{11}{12}$ & &
$\frac{35}{36}$ \\
\bottomrule
\end{tabular}}}
\end{minipage}}
\caption{Metrical hierarchy from Ockeghem's Missa prolationem expressed as $F_{36}'$}
\label{tab:1}
\end{table}

The hierarchical pattern of one of Ockeghem's voices in Table
\ref{tab:1} shows also some interesting properties that lead to a
general method for the construction of such tables. Starting with the
first metrical level (I) in Table \ref{tab:1}, we see that the largest
denominator is always equal to the number of measures required to come
back to square one with all the other voices, i.e the number of
measures per hyper-meter. All the other denominators on level I are
divisors of the largest denominator. The next level (II) shows us how
many subdivisions are contained in the hyper-meter on that particular
level. This is again indicated by the largest denominator of the level
and all other denominators are his divisors \emph{excluding} those
already contained on all lower indexed levels. This principle repeats
itself on all higher indexed metrical levels. For each denominator $n$
in the scheme, the numerators also follow a simple principle: they
traverse the complete ordered list of numbers co-prime to $n$.

As we have shown in \citeN{Farey} and \citeN{gb:husserl}, the Farey
Sequence is ideal for tasks like rhythmic modelling, music performance
analysis and music theory.
Chapter 3 of \citeN{hardy} gives a description and proves the properties of
the Farey Sequence. Its scalability and general independence from the
concepts of bars and meter is of advantage because it can be applied
to numerous different musical styles.  We have given a glimpse in the
above Renaissance example (Table \ref{tab:1}) how it could be used to
encode polyrhythmic structures.  Other examples can include African
Polyrhythm, Western Classical, Avantgarde and Popular Music, Greek
Verse Rhythms, Indian Percussion and many more. The principle remains
always the same. 

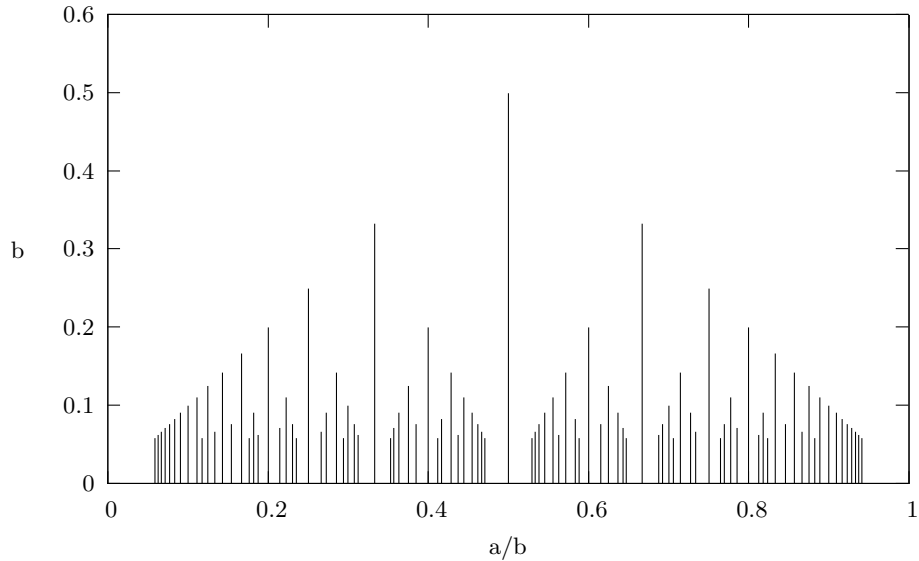
\begin {figure}
  \begin{center}
\setlength{\unitlength}{0.240900pt}
\ifx\plotpoint\undefined\newsavebox{\plotpoint}\fi
\sbox{\plotpoint}{\rule[-0.200pt]{0.400pt}{0.400pt}}%
\begin{picture}(1500,900)(0,0)
\sbox{\plotpoint}{\rule[-0.200pt]{0.400pt}{0.400pt}}%
\put(181.0,123.0){\rule[-0.200pt]{4.818pt}{0.400pt}}
\put(161,123){\makebox(0,0)[r]{ 0}}
\put(1419.0,123.0){\rule[-0.200pt]{4.818pt}{0.400pt}}
\put(181.0,246.0){\rule[-0.200pt]{4.818pt}{0.400pt}}
\put(161,246){\makebox(0,0)[r]{ 0.1}}
\put(1419.0,246.0){\rule[-0.200pt]{4.818pt}{0.400pt}}
\put(181.0,369.0){\rule[-0.200pt]{4.818pt}{0.400pt}}
\put(161,369){\makebox(0,0)[r]{ 0.2}}
\put(1419.0,369.0){\rule[-0.200pt]{4.818pt}{0.400pt}}
\put(181.0,492.0){\rule[-0.200pt]{4.818pt}{0.400pt}}
\put(161,492){\makebox(0,0)[r]{ 0.3}}
\put(1419.0,492.0){\rule[-0.200pt]{4.818pt}{0.400pt}}
\put(181.0,614.0){\rule[-0.200pt]{4.818pt}{0.400pt}}
\put(161,614){\makebox(0,0)[r]{ 0.4}}
\put(1419.0,614.0){\rule[-0.200pt]{4.818pt}{0.400pt}}
\put(181.0,737.0){\rule[-0.200pt]{4.818pt}{0.400pt}}
\put(161,737){\makebox(0,0)[r]{ 0.5}}
\put(1419.0,737.0){\rule[-0.200pt]{4.818pt}{0.400pt}}
\put(181.0,860.0){\rule[-0.200pt]{4.818pt}{0.400pt}}
\put(161,860){\makebox(0,0)[r]{ 0.6}}
\put(1419.0,860.0){\rule[-0.200pt]{4.818pt}{0.400pt}}
\put(181.0,123.0){\rule[-0.200pt]{0.400pt}{4.818pt}}
\put(181,82){\makebox(0,0){ 0}}
\put(181.0,840.0){\rule[-0.200pt]{0.400pt}{4.818pt}}
\put(433.0,123.0){\rule[-0.200pt]{0.400pt}{4.818pt}}
\put(433,82){\makebox(0,0){ 0.2}}
\put(433.0,840.0){\rule[-0.200pt]{0.400pt}{4.818pt}}
\put(684.0,123.0){\rule[-0.200pt]{0.400pt}{4.818pt}}
\put(684,82){\makebox(0,0){ 0.4}}
\put(684.0,840.0){\rule[-0.200pt]{0.400pt}{4.818pt}}
\put(936.0,123.0){\rule[-0.200pt]{0.400pt}{4.818pt}}
\put(936,82){\makebox(0,0){ 0.6}}
\put(936.0,840.0){\rule[-0.200pt]{0.400pt}{4.818pt}}
\put(1187.0,123.0){\rule[-0.200pt]{0.400pt}{4.818pt}}
\put(1187,82){\makebox(0,0){ 0.8}}
\put(1187.0,840.0){\rule[-0.200pt]{0.400pt}{4.818pt}}
\put(1439.0,123.0){\rule[-0.200pt]{0.400pt}{4.818pt}}
\put(1439,82){\makebox(0,0){ 1}}
\put(1439.0,840.0){\rule[-0.200pt]{0.400pt}{4.818pt}}
\put(181.0,123.0){\rule[-0.200pt]{303.052pt}{0.400pt}}
\put(1439.0,123.0){\rule[-0.200pt]{0.400pt}{177.543pt}}
\put(181.0,860.0){\rule[-0.200pt]{303.052pt}{0.400pt}}
\put(181.0,123.0){\rule[-0.200pt]{0.400pt}{177.543pt}}
\put(40,491){\makebox(0,0){b}}
\put(810,21){\makebox(0,0){a/b}}
\put(255.0,123.0){\rule[-0.200pt]{0.400pt}{17.345pt}}
\put(260.0,123.0){\rule[-0.200pt]{0.400pt}{18.549pt}}
\put(265.0,123.0){\rule[-0.200pt]{0.400pt}{19.754pt}}
\put(271.0,123.0){\rule[-0.200pt]{0.400pt}{21.199pt}}
\put(278.0,123.0){\rule[-0.200pt]{0.400pt}{22.645pt}}
\put(286.0,123.0){\rule[-0.200pt]{0.400pt}{24.572pt}}
\put(295.0,123.0){\rule[-0.200pt]{0.400pt}{26.981pt}}
\put(307.0,123.0){\rule[-0.200pt]{0.400pt}{29.631pt}}
\put(321.0,123.0){\rule[-0.200pt]{0.400pt}{32.762pt}}
\put(329.0,123.0){\rule[-0.200pt]{0.400pt}{17.345pt}}
\put(338.0,123.0){\rule[-0.200pt]{0.400pt}{37.099pt}}
\put(349.0,123.0){\rule[-0.200pt]{0.400pt}{19.754pt}}
\put(361.0,123.0){\rule[-0.200pt]{0.400pt}{42.157pt}}
\put(375.0,123.0){\rule[-0.200pt]{0.400pt}{22.645pt}}
\put(391.0,123.0){\rule[-0.200pt]{0.400pt}{49.384pt}}
\put(403.0,123.0){\rule[-0.200pt]{0.400pt}{17.345pt}}
\put(410.0,123.0){\rule[-0.200pt]{0.400pt}{26.981pt}}
\put(417.0,123.0){\rule[-0.200pt]{0.400pt}{18.549pt}}
\put(433.0,123.0){\rule[-0.200pt]{0.400pt}{59.261pt}}
\put(451.0,123.0){\rule[-0.200pt]{0.400pt}{21.199pt}}
\put(461.0,123.0){\rule[-0.200pt]{0.400pt}{32.762pt}}
\put(471.0,123.0){\rule[-0.200pt]{0.400pt}{22.645pt}}
\put(477.0,123.0){\rule[-0.200pt]{0.400pt}{17.345pt}}
\put(496.0,123.0){\rule[-0.200pt]{0.400pt}{73.956pt}}
\put(516.0,123.0){\rule[-0.200pt]{0.400pt}{19.754pt}}
\put(524.0,123.0){\rule[-0.200pt]{0.400pt}{26.981pt}}
\put(540.0,123.0){\rule[-0.200pt]{0.400pt}{42.157pt}}
\put(551.0,123.0){\rule[-0.200pt]{0.400pt}{17.345pt}}
\put(558.0,123.0){\rule[-0.200pt]{0.400pt}{29.631pt}}
\put(568.0,123.0){\rule[-0.200pt]{0.400pt}{22.645pt}}
\put(574.0,123.0){\rule[-0.200pt]{0.400pt}{18.549pt}}
\put(600.0,123.0){\rule[-0.200pt]{0.400pt}{98.528pt}}
\put(625.0,123.0){\rule[-0.200pt]{0.400pt}{17.345pt}}
\put(630.0,123.0){\rule[-0.200pt]{0.400pt}{21.199pt}}
\put(638.0,123.0){\rule[-0.200pt]{0.400pt}{26.981pt}}
\put(653.0,123.0){\rule[-0.200pt]{0.400pt}{37.099pt}}
\put(665.0,123.0){\rule[-0.200pt]{0.400pt}{22.645pt}}
\put(684.0,123.0){\rule[-0.200pt]{0.400pt}{59.261pt}}
\put(699.0,123.0){\rule[-0.200pt]{0.400pt}{17.345pt}}
\put(705.0,123.0){\rule[-0.200pt]{0.400pt}{24.572pt}}
\put(720.0,123.0){\rule[-0.200pt]{0.400pt}{42.157pt}}
\put(731.0,123.0){\rule[-0.200pt]{0.400pt}{18.549pt}}
\put(740.0,123.0){\rule[-0.200pt]{0.400pt}{32.762pt}}
\put(753.0,123.0){\rule[-0.200pt]{0.400pt}{26.981pt}}
\put(762.0,123.0){\rule[-0.200pt]{0.400pt}{22.645pt}}
\put(768.0,123.0){\rule[-0.200pt]{0.400pt}{19.754pt}}
\put(773.0,123.0){\rule[-0.200pt]{0.400pt}{17.345pt}}
\put(810.0,123.0){\rule[-0.200pt]{0.400pt}{147.913pt}}
\put(847.0,123.0){\rule[-0.200pt]{0.400pt}{17.345pt}}
\put(852.0,123.0){\rule[-0.200pt]{0.400pt}{19.754pt}}
\put(858.0,123.0){\rule[-0.200pt]{0.400pt}{22.645pt}}
\put(867.0,123.0){\rule[-0.200pt]{0.400pt}{26.981pt}}
\put(880.0,123.0){\rule[-0.200pt]{0.400pt}{32.762pt}}
\put(889.0,123.0){\rule[-0.200pt]{0.400pt}{18.549pt}}
\put(900.0,123.0){\rule[-0.200pt]{0.400pt}{42.157pt}}
\put(915.0,123.0){\rule[-0.200pt]{0.400pt}{24.572pt}}
\put(921.0,123.0){\rule[-0.200pt]{0.400pt}{17.345pt}}
\put(936.0,123.0){\rule[-0.200pt]{0.400pt}{59.261pt}}
\put(955.0,123.0){\rule[-0.200pt]{0.400pt}{22.645pt}}
\put(967.0,123.0){\rule[-0.200pt]{0.400pt}{37.099pt}}
\put(982.0,123.0){\rule[-0.200pt]{0.400pt}{26.981pt}}
\put(990.0,123.0){\rule[-0.200pt]{0.400pt}{21.199pt}}
\put(995.0,123.0){\rule[-0.200pt]{0.400pt}{17.345pt}}
\put(1020.0,123.0){\rule[-0.200pt]{0.400pt}{98.528pt}}
\put(1046.0,123.0){\rule[-0.200pt]{0.400pt}{18.549pt}}
\put(1052.0,123.0){\rule[-0.200pt]{0.400pt}{22.645pt}}
\put(1062.0,123.0){\rule[-0.200pt]{0.400pt}{29.631pt}}
\put(1069.0,123.0){\rule[-0.200pt]{0.400pt}{17.345pt}}
\put(1080.0,123.0){\rule[-0.200pt]{0.400pt}{42.157pt}}
\put(1096.0,123.0){\rule[-0.200pt]{0.400pt}{26.981pt}}
\put(1104.0,123.0){\rule[-0.200pt]{0.400pt}{19.754pt}}
\put(1125.0,123.0){\rule[-0.200pt]{0.400pt}{73.956pt}}
\put(1143.0,123.0){\rule[-0.200pt]{0.400pt}{17.345pt}}
\put(1149.0,123.0){\rule[-0.200pt]{0.400pt}{22.645pt}}
\put(1159.0,123.0){\rule[-0.200pt]{0.400pt}{32.762pt}}
\put(1169.0,123.0){\rule[-0.200pt]{0.400pt}{21.199pt}}
\put(1187.0,123.0){\rule[-0.200pt]{0.400pt}{59.261pt}}
\put(1203.0,123.0){\rule[-0.200pt]{0.400pt}{18.549pt}}
\put(1210.0,123.0){\rule[-0.200pt]{0.400pt}{26.981pt}}
\put(1217.0,123.0){\rule[-0.200pt]{0.400pt}{17.345pt}}
\put(1229.0,123.0){\rule[-0.200pt]{0.400pt}{49.384pt}}
\put(1245.0,123.0){\rule[-0.200pt]{0.400pt}{22.645pt}}
\put(1259.0,123.0){\rule[-0.200pt]{0.400pt}{42.157pt}}
\put(1271.0,123.0){\rule[-0.200pt]{0.400pt}{19.754pt}}
\put(1282.0,123.0){\rule[-0.200pt]{0.400pt}{37.099pt}}
\put(1291.0,123.0){\rule[-0.200pt]{0.400pt}{17.345pt}}
\put(1299.0,123.0){\rule[-0.200pt]{0.400pt}{32.762pt}}
\put(1313.0,123.0){\rule[-0.200pt]{0.400pt}{29.631pt}}
\put(1325.0,123.0){\rule[-0.200pt]{0.400pt}{26.981pt}}
\put(1334.0,123.0){\rule[-0.200pt]{0.400pt}{24.572pt}}
\put(1342.0,123.0){\rule[-0.200pt]{0.400pt}{22.645pt}}
\put(1349.0,123.0){\rule[-0.200pt]{0.400pt}{21.199pt}}
\put(1355.0,123.0){\rule[-0.200pt]{0.400pt}{19.754pt}}
\put(1360.0,123.0){\rule[-0.200pt]{0.400pt}{18.549pt}}
\put(1365.0,123.0){\rule[-0.200pt]{0.400pt}{17.345pt}}
\put(181.0,123.0){\rule[-0.200pt]{303.052pt}{0.400pt}}
\put(1439.0,123.0){\rule[-0.200pt]{0.400pt}{177.543pt}}
\put(181.0,860.0){\rule[-0.200pt]{303.052pt}{0.400pt}}
\put(181.0,123.0){\rule[-0.200pt]{0.400pt}{177.543pt}}
\end{picture}
  \end{center}
  \caption[$F_{17}$]{Correlation of $a/b \in F_{17}$ and $1/b$ in the
    interval $(0, 1)$}
  \label{fig:farey17}
\end {figure}

The visualisation in Figure~\ref{fig:farey17} clearly
points out that the smaller the denominator the larger are the
symmetrical gaps around the x-position of the ratio, {\em i.e.\/} the smaller
$b$ of the ratio $a/b$, the greater the distance.  We believe that it is
due to those relatively large gaps surrounding simpler ratios that
they form perceptually useful zones of attraction for more complex
ratios that fall into them or that come close enough. It is left for
future field studies to measure and to find evidence whether these
zones of attraction are perceptually relevant or not. Composers who
want to ``stay away'' from those simple ratios will need to leave a
considerable amount of space around these zones of attraction. 
It becomes also clear that there are various accelerandi and
ritardandi encoded in every $F_n$, for example there are
gestalts\footnote{That is the concept of shape or configuration of a
  whole perceived object} that
form visible triangles between larger reciprocals and smaller ones in
their surrounding area (Figure \ref{fig:farey17}). These gestalts are
formed by monotonically increasing or decreasing values of the
denominators. The increasing or decreasing tendencies overlap each
other. The gaps between the ratios forming those triangles are always
on a logarithmic scale, hence the impression of accelerated or reduced
tempo that becomes evident through the sonification of these
triangular gestalts. It is well known that timed durations need to be
placed on a logarithmic rather than linear scale in order to convey a
``natural'' sense of spacing and tempo modification. These structures
are clearly mirrored around the $1/2$ value that is part of every
$F_n$ with $n>1$.

The great variety of styles and concepts from medieval to modern eras
can be realised by applying intelligent filtering methods to sieve
through the sequences. Examples for filtering include probabilistic
methods and filters exploiting the prime number composition of
integers and ratios, for example b-smooth numbers, or Clarence
Barlow's function for the `Indigestibility' of a natural integer
\cite{hajdu}. The Farey Sequence has been known for a while in the
area of musical tuning systems\footnote{For example Erv Wilson's
  annotations of tunings used by \citeN{partch1979}
  http://www.anaphoria.com/wilson.html}. Its use for rhythmic
modelling has not been fully exploited yet; \systemname\ is the first
music application for composition where rhythms and musical forms will
be generated on the basis of the principles outlined in this
paper. The Farey Sequence $F_n$ is per se highly symmetrical and
unfolds harmonic subdivisions of unity via a recursive calculation of
mediant fractions\footnote{See
  http://mathworld.wolfram.com/FareySequence.html}. Every music that
depends on an underlying beat or pulsation can be represented by using
$F_n$ to denote the normalised occurrence of musical events, e.g. note
onsets.

Off-Beat rhythms are extremely useful for generative purposes. Two or
more of these rhythms stacked in layers can always 
generate new combinations by using different accentuation patterns and
different dynamic processes. Again, the Farey-Sequence proves to be a
very useful structure to realise this concept.  The details are beyond
what is necessary in this paper, but in \citeN{Boenn09} it is shown
that many styles like Bebop, Funk and 20th century Avantgarde, are
modelled by this mechanism.  Speech rhythm, as used in various musical
styles are also within the scope.

Finally, Sima Arom's study \cite{sa:apap} on African Polyrhythms has
been very influential on contemporary western composers because of his
successful recording and transcription processes that form the basis
of his further analysis. We are seeking to translate some of these
principles into features for \systemname\ for
creative purposes but also in order to proof the general use of our
concept.  But this is for the future (Section~\ref{future}).

The main message is that the Farey sequence contains all that is
necessary for a very wide range of rhythmic patterns.  With an
encoding of them in \AnsProlog{} we have all the
infrastructure we need.

\subsection{Rhythm in \systemname}
\label{FareyProg}

The question now arises how we can combine this generative model for
rhythm with the rules for melody, counterpoint and harmony that have
been already implemented in \systemname. Of central importance for
musical experiences in our view is the constant inter-change of
\emph{impact} and \emph{resolution} that influences the behaviour of
musical parameters on micro- and macro-structural levels. One can
compare \emph{impact} with gathering musical energy \cite{Thakar} and
\emph{resolution} with the release of previously built-up energy. We
were very careful to make sure that the melodic lines generated are
following this principle. With the encoding of rhythm we have now the
possibility to precisely control the timing aspect of when to turn the
musical movement from impact to resolution.

Rhythm in \systemnamev{1.5} is the first prototype and an improved version
is under development.
As mentioned earlier, the encoding of rhythm is based on the concept
of Farey sequences.  One can view each rational number in the sequence
as a way of partitioning the range [0,1], and as any particular
partitioning can be constructed from a series of partitioning by a
rational number with a prime denominator, the full collection of
possible rhythms can be considered as all possible trees of
partitions.   In the initial version we limit the partition
denominators to be taken to be only the primes 2 and 3, which allows
a sufficiently rich subset of trees (and hence rhythms) for our
musical genre.  The encoding allows other primes, at some
computational cost.

\begin{figure}[tbp]
\hspace{-1cm}
\scalebox{0.7}{
\begin{ttfamily}
\begin{tabular}{l}
\%\% Each Farey tree has a given depth\\
depth(F,MD + BD + DD) :- measureDepth(MD), beatDepth(F,BD), durationDepth(F,DD).\\
level(F,1..DE) :- depth(F,DE).\\
\mbox{}\\

\%\% Each Farey tree is divided into three layers (top to bottom)\\
\%\% Measure, beats and note duration\\
\%\% (bars, time signature and note value)\\
measureLevel(F,FL) :- depth(F,DE), durationDepth(F,DD), beatDepth(F,BD), \\
\mbox{\hspace{3cm}} level(F,FL), FL <= DE - (DD + BD).\\
measureLeafLevel(F,DE - (DD + BD)) :- depth(F,DE), durationDepth(F,DD), beatDepth(F,BD).\\
\mbox{}\\

\%\% Map from nodes to time positions\\
\%\% Mapping increments each time a node is present\\
nodeStep(F,0,1).\\
nodeStep(F,ND,T) :- not present(F,DLL,ND), nodeStep(F,ND-1,T),\\
 \mbox{\hspace{3cm}}  node(F,DLL,ND), durationLeafLevel(F,DLL), ND > 0.\\
nodeStep(F,ND,T+1) :-   present(F,DLL,ND), nodeStep(F,ND-1,T),\\
 \mbox{\hspace{3cm}}  node(F,DLL,ND), durationLeafLevel(F,DLL), ND > 0.\\
\%\% From this we derive a unique mapping from node to time step\\
timeToNode(P,1,0).\\
timeToNode(P,T,ND) :- present(F,DLL,ND), nodeStep(F,ND-1,T-1),\\
 \mbox{\hspace{3cm}}  node(F,DLL,ND), durationLeafLevel(F,DLL), ND > 0,\\
 \mbox{\hspace{3cm}}  partToFareyTree(P,F).\\
\mbox{}\\

\%\% Beat strength is created at the first level of the beat layer ...\\
nodeBeatStrength(F,MLL+1,ND2,1) :- measureLeafLevel(F,MLL), node(F,MLL,ND1),\\
 \mbox{\hspace{3cm}}  descendant(F,0,MLL,ND1,MLL+1,ND2).\\
nodeBeatStrength(F,MLL+1,ND2,0) :- measureLeafLevel(F,MLL), node(F,MLL,ND1),\\
 \mbox{\hspace{3cm}}  descendant(F,D,MLL,ND1,MLL+1,ND2), D != 0.\\
\mbox{}\\

\%\% Lowest and highest notes must also be slower\\
playsHeighestNote(P,T) :- chosenNote(P,T,N), lowestNote(P,N).\\
playsLowestNote(P,T) :- chosenNote(P,T,N), lowestNote(P,N).\\
\ :- playsHeighestNote(P,T), timeStepDuration(P,T,DS), DS > 1.\\
\ :- playsLowestNote(P,T), timeStepDuration(P,T,DS), DS > 1.\\
\end{tabular}
\end{ttfamily}
}

\caption{A small rhythm code fragment}\label{code:rhythm}
\end{figure}

\begin{figure*}[t]
\begin{center}
\includegraphics[width=0.8\textwidth]{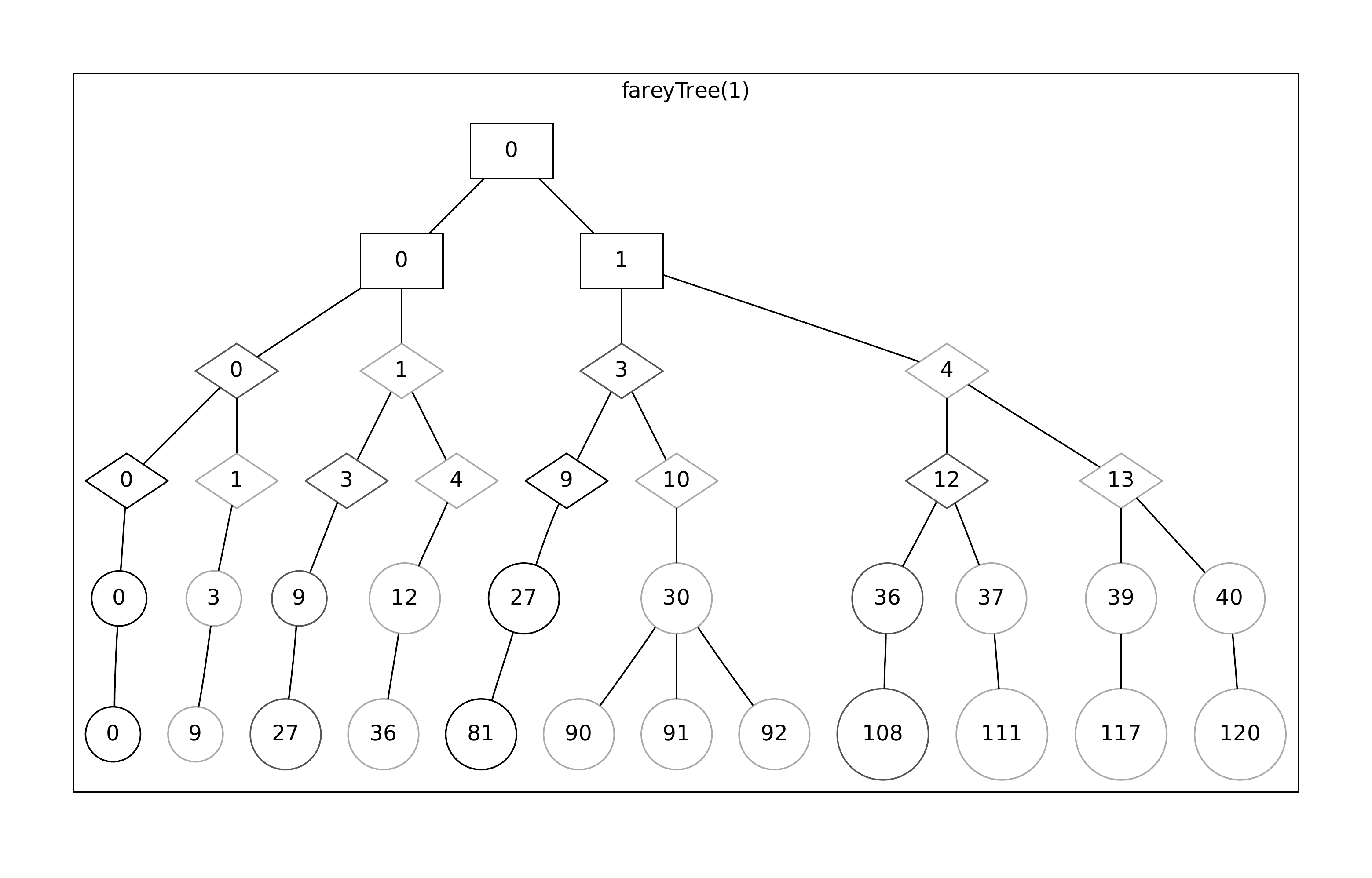}

\begin{verbatim}
             | 65 67 65 77 76 69 71 76 72 74 76 77 
             |  F  G  F  F' E' A  B  E' C' D' E' F'
             |   +2 -2 +12 -1 -7 +2 +5 -4 +2 +2 +1 
             | (((X X) (X X)) ((X (X X X)) ((X X) (X X))))

             | 65 60 62 62 60 65 67 67 69 67 67 65 
             |  F  C  D  D  C  F  G  G  A  G  G  F 
             |  -5 +2 "" -2 +5 +2 "" +2 -2 "" -2 
             | (((X X) (X X)) ((X (X X X)) ((X X) (X X))))
\end{verbatim}
\end{center}
\caption{The partitioning tree and human readable format for the Lydian duet composition.}\label{score2}
\end{figure*}

While the Farey sequence is a useful form for traversing the rhythmic structure, the partitioning tree is a useful (computational) way of building up a filtered Farey sequence by using a top-down approach, starting with a section of time and working it down by way of making subdivisions. The filtering of the Farey sequence is then achieved through the constraint of allowing only small primes as subdividing numbers within the tree, i.e. 2 and 3 in our initial implementation of rhythm. 
The further advantage of having a tree is that it allows us to have an easy access to different metrical levels (measure, beat, notes), which is vital for the later addition of rules about impact/resolution and for rules governing usage of consonance/dissonance. An example of a partitioning tree can be found in Figure~\ref{score2}.  
Each part is divided in a number of measures, these forms the top layer of the tree (represented by rectangles in the example below). Each measure is then divided in a number of beats (diamond shapes), which control the emphasis of notes (the weight of shading).The notes (circles) of the part are then grouped w.r.t. duration and placed within their respective beats. Individual notes are the leaves of the tree. 

The current version, \systemnamev{1.5}, picks an arbitrary tree from a constrained set of
possibilities and imposes it on the output, so all
parts have the same rhythm. In future versions, 
melodic and harmonic changes will interact with the decision
over which tree is chosen.
The newest version, which is not yet
released, extends this by adding rules to deal with beats, their
strength, duration and the interplay between parts.   These rules
allow the parts to have independent, but related rhythms.

Figure~\ref{code:rhythm} contains a small code fragment of the current
rhythm section of the system.  The fragments show a portion of the
structure of the partition tree and its three layers of measures, beats and notes.  
Each of these layers have
separate rules that govern them.  The fragment also shows the 
mapping from notes to time instances.  We have included the definition
of beat strength and one of the constraints linking tone and speed.

While not shown in this code fragment, the rhythm code is, like the
other sections, written in such a way that diagnosis is possible.

In order to include rhythm in a composition, it suffices the include
option \texttt{--rhythm} to the \filename{programBuilder.pl} script. 
The command:

{\small
\begin{verbatim}
./programBuilder.pl --task=compose --mode=lydian --time=12 --style=duet  
--rhythm > rhythm-comp 
\end{verbatim}
}

\noindent composes a duet in Lydian with 12 notes per part.

The program can then be parsed using \filename{parse.lp} to generate
\prog{Csound}, \prog{Lilypond} and human readable output.
Figure~\ref{score2} shows the partitioning tree and the human score for the Lydian duet.

\section{Evaluation}\label{results}

In this section we evaluate the performance of \systemname{} both from a computational 
 and a musical perspective. We mainly focus on the non-rhythm part of the released 
version 1.5 for stability and reproducibility.

\subsection{Run-time Results}

To evaluate the practicality of using answer set programming in a
composition system we timed \systemnamev{1.5} without the rhythm version 
while composing a suite of score with increasing difficulty.

Tables \ref{timing:solo}-\ref{timing:quartet} contains the timings for
a number of answer set solvers (\prog{clasp} \cite{gekanesc07a},
\prog{cmodels} \cite{cmodels}, \prog{smodels} \cite{smodels},
\prog{smodels-ie} \cite{smodels-ie}, \prog{smodelscc}
\cite{smodelscc}, and \prog{sup} \cite{sup}) in composing solos,
duets, trios and quartets of a given length.

\begin{table}[tbp]{\tiny{\begin{tabular}[t]{ccccccc}
\toprule 
\mbox{\hspace{27pt}} & \multicolumn{6}{c}{Solvers} \\ \cmidrule(lr){2-7}Length \  & Clasp 1.2.1\ \ & Cmodels 3.79\ \ & Smodels 2.33\ \ & Smodels-IE 1.0.0\ \ & Smodels\_cc 1.08\ \ & Sup 0.4\ \ \\ 
\midrule
4 & 0.02 & 0.09 & 0.04 & 0.02 & 0.10 & 0.04 \\
8 & 0.18 & 0.38 & 1.27 & 0.52 & 4.55 & 0.16 \\
12 & 0.48 & 1.10 & 8.99 & 2.87 & 27.45 & 0.64 \\
16 & 1.05 & 2.06 & 36.03 & 10.19 & 86.56 & 1.01 \\
20 & 2.68 & 3.11 & 32.52 & 10.02 & 93.61 & 2.06 \\
24 & 2.42 & 4.22 & 193.40 & 58.06 & Time out & 2.11 \\
28 & 3.84 & 5.80 & 239.49 & 80.56 & Time out & 4.02 \\
32 & 3.90 & 7.11 & 305.05 & 102.91 & Time out & 4.66 \\
\bottomrule
\end{tabular}}}\caption{Time taken (in seconds) for a number of solvers generating a solo piece.}\label{timing:solo}\end{table}

\begin{table}[tbp]{\tiny{\begin{tabular}[t]{ccccccc}
\toprule 
\mbox{\hspace{27pt}} & \multicolumn{6}{c}{Solvers} \\ \cmidrule(lr){2-7}Length \  & Clasp 1.2.1\ \ & Cmodels 3.79\ \ & Smodels 2.33\ \ & Smodels-IE 1.0.0\ \ & Smodels\_cc 1.08\ \ & Sup 0.4\ \ \\ 
\midrule
4 & 0.14 & 0.28 & 0.23 & 0.10 & 0.64 & 0.14 \\
8 & 0.43 & 0.98 & 10.60 & 4.92 & 77.07 & 0.57 \\
12 & 2.28 & 2.54 & Time out & Time out & Time out & 2.18 \\
16 & 1.91 & 3.92 & Time out & Time out & Time out & 3.33 \\
20 & 3.12 & 6.58 & Time out & Time out & Time out & 7.86 \\
24 & 8.60 & 8.71 & Time out & Time out & Time out & 13.55 \\
28 & 14.94 & 19.29 & Time out & Time out & Time out & 31.05 \\
32 & 13.56 & 26.90 & Time out & Time out & Time out & 31.63 \\
\bottomrule
\end{tabular}}}\caption{Time taken (in seconds) for a number of solvers generating a duet piece.}\label{timing:duet}\end{table}

\begin{table}[tbp]{\tiny{\begin{tabular}[t]{ccccccc}
\toprule 
\mbox{\hspace{27pt}} & \multicolumn{6}{c}{Solvers} \\ \cmidrule(lr){2-7}Length \  & Clasp 1.2.1\ \ & Cmodels 3.79\ \ & Smodels 2.33\ \ & Smodels-IE 1.0.0\ \ & Smodels\_cc 1.08\ \ & Sup 0.4\ \ \\ 
\midrule
4 & 0.26 & 0.30 & 0.27 & 0.04 & 0.30 & 0.23 \\
8 & 0.78 & 1.52 & 3.42 & 1.34 & 10.39 & 0.96 \\
12 & 2.17 & 3.20 & 22.44 & 8.52 & 77.80 & 2.88 \\
16 & 2.81 & 5.35 & 104.68 & 39.20 & Time out & 4.55 \\
20 & 6.95 & 7.63 & Time out & Time out & Time out & 9.01 \\
24 & 11.90 & 10.92 & Time out & Time out & Time out & 9.75 \\
28 & 41.28 & 12.32 & Time out & Time out & Time out & 10.68 \\
32 & 52.98 & 19.33 & Time out & Time out & Time out & 47.94 \\
\bottomrule
\end{tabular}}}\caption{Time taken (in seconds) for a number of solvers generating a trio piece.}\label{timing:trio}\end{table}

\begin{table}[tbp]{\tiny{\begin{tabular}[t]{ccccccc}
\toprule 
\mbox{\hspace{27pt}} & \multicolumn{6}{c}{Solvers} \\ \cmidrule(lr){2-7}Length \  & Clasp 1.2.1\ \ & Cmodels 3.79\ \ & Smodels 2.33\ \ & Smodels-IE 1.0.0\ \ & Smodels\_cc 1.08\ \ & Sup 0.4\ \ \\ 
\midrule
4 & 0.43 & 0.55 & 0.49 & 0.08 & 0.63 & 0.55 \\
8 & 1.14 & 2.55 & 7.98 & 3.28 & 32.20 & 2.70 \\
12 & 4.24 & 5.38 & 35.11 & 13.07 & 128.75 & 5.46 \\
16 & 9.59 & 8.64 & 336.36 & 126.21 & Time out & 16.86 \\
20 & 69.37 & 11.87 & Time out & Time out & Time out & 17.44 \\
24 & 194.73 & 20.44 & Time out & Time out & Time out & 30.99 \\
28 & 246.10 & 19.32 & Time out & Time out & Time out & 79.13 \\
32 & Time out & 46.61 & Time out & Time out & Time out & 113.30 \\
\bottomrule
\end{tabular}}}\caption{Time taken (in seconds) for a number of solvers generating a quartet piece.}\label{timing:quartet}\end{table}

The experiments were run using a 2.4Ghz AMD Athlon X2 4600+ processor,
running a 64 bit version of OpenSuSE 11.1.  All solvers were built in
32 bit mode.  Each test was limited to 10 minutes of CPU time and 2Gb
of RAM.  Programs were ground using \prog{GrinGo} 2.0.3 and grounding
times of between 0.5 and 7 seconds were excluded. They were not
reported for two distinct reasons: first, the grounder times are
constants as we used the same grounder for each solver and second, grounding is set-up time. 
Just like in a life concert all the equipment has
already been set before the concert, so should grounding be considered separately from the real-time
composition. 
All solvers were run using default options, except
\prog{cmodels} which was set to use the \prog{MiniSAT} 2.0 back end as
opposed to the default (\prog{zchaff}). The programs used are
available from \url{http://www.cs.bath.ac.uk/~mjb/anton}.

The results show a significant increase in performance from 
\systemnamev{1.0} reported in \cite{bbdf}.  Then we
were only able to compose duets up to length 16, which took 29.63
seconds using the fastest solver \prog{clasp}. The current system,
\systemnamev{1.5}, only takes 1.01 seconds for the same composition
using the same solver. Furthermore, we can now compose trios and
quartets within a reasonable time frame.

While improvements in the underlying solvers are a factor in the
steep increase in performance, they are not the main contributor.  We
obtained most of our increases by revisiting \AnsProlog encoding and finding
more compact representations. We have compacted the rule set of minor keys
which gives some reduction in space and run time. We reformatted some
of the harmony rules and relaxed them so they only apply to
neighbouring parts rather than all parts.  The removal of redundant
constraints compacts the program by a surprising amount.  The
rewriting of the repeated notes section produced a massive increase in
grounding while the improved encoding of highest and lowest note saved
us about 150,000 grounded rules on 16 note duet and about 30\% in run
time.  Ranges over two octaves is now only noted at the end of the
program, rather than at the point at which it is triggered.  While
this is slightly less informative, it offers a more than significant
speed-up.

These results show that the system, using the more powerful
solvers, is not only fast enough to be used as a component in an
interactive composition tool but, when restricting to shorter
sequences, could be used for real-time generation of music.

It is also interesting to note that the only solvers able to generate longer
sequences using two or more parts all implement clause learning strategies, suggesting
that the problem is particularly susceptible to this kind of technique.

We have not included run-time results for the rhythm section. The
implementation as this feature is still being explored and
the results would not be representative.

\subsection{Music Quality}

\begin{figure}[bth]
\begin{center}
\includegraphics[width=0.9\textwidth]{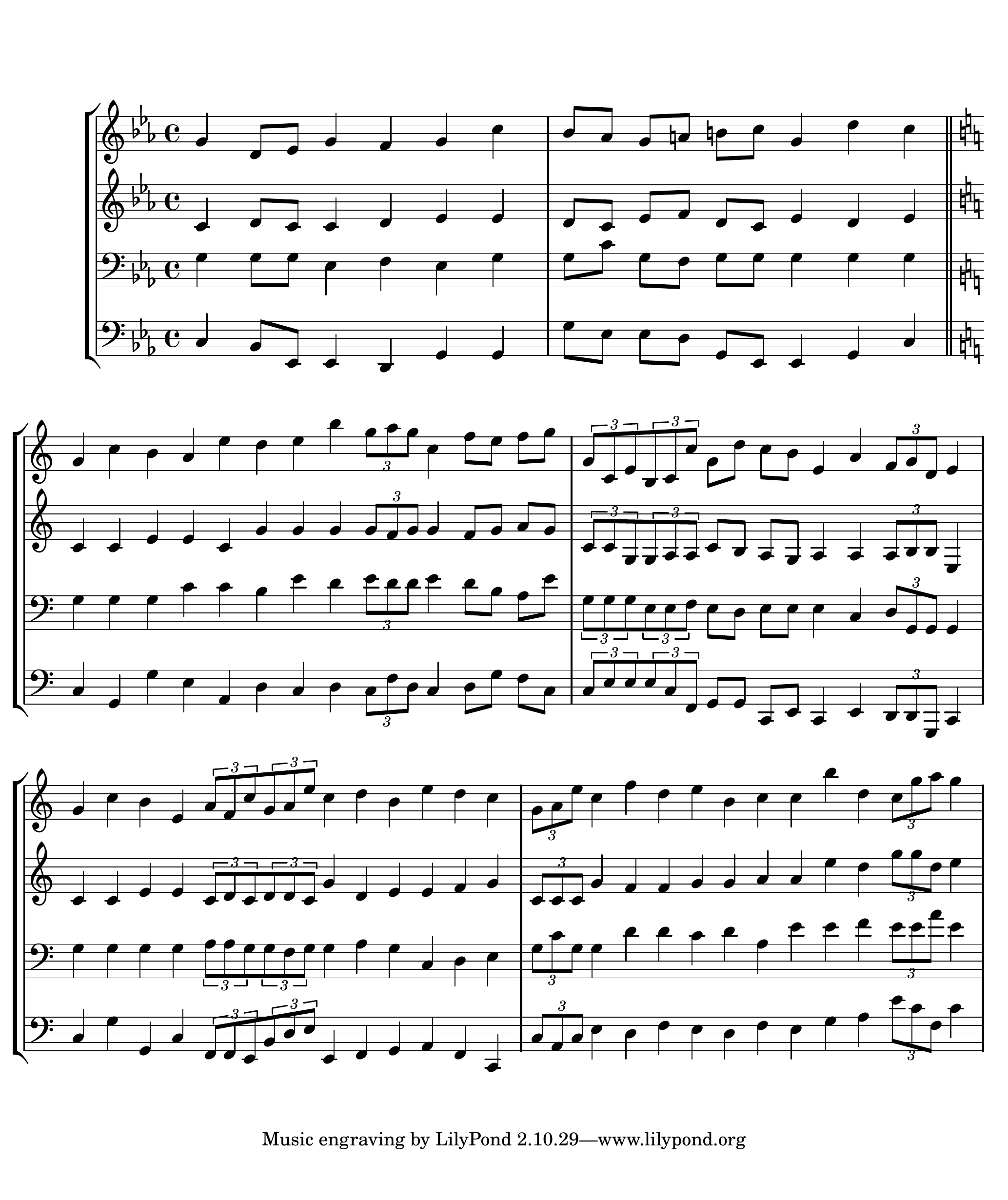}
\end{center}
\vspace{-10pt}
The interested reader can find more examples on the web:
\url{http://dream.cs.bath.ac.uk/Anton}
\caption{Fragments by \systemname}
\label{fragments}
\end{figure}

The other way to evaluate the system is to judge the music it
produces.  This is a less certain process, involving personal values.
However we feel that the music is acceptable, at least of the quality
of some students of composition, and at times capable of moments of
excitement. Pieces by \systemnamev{1.0} have been played to a number of
musicians, who apart from the rhythmic deficiency we are addressing
have agreed that it is valid music.  The introduction of rhythm is
more recent, and consequently it has not been subjected to so much
scrutiny.  There are still some refinements that could improve the
output
but many of the short pieces are clearly valid, and musical.  In
figure~\ref{fragments} we present a short quartet sequence in the
minor key, followed by four major key pieces composed using the \systemname{} 
system;
the audio and score can be found
in the same location as the other works.

\section{The Use of ASP in \systemname}\label{asp-kr}

\subsection{Why ASP?}

While music appreciation is matter of personal taste, musicologists
use sets of rules which determine to which style a musical composition belongs
or whether a piece breaks or expands the common practice of a certain
composer or era. These sets of rules also govern the composition.  So an intuitive
and obvious approach to automatic composition is to encode these rules and
use a rule based algorithm to produce valid music compositions. This
natural and simple way of encoding things is shown in terms of speed of
development, roughly 2 man-months, sophistication of the results, the
amount of code (about 500 lines of code) and flexibility; we can not
only easily encode different styles but the same application is usable
not only
for automated composition but also diagnosis and human assisted
composition. Furthermore, we automatically gain from all improvements
in the underlying solver.

\subsection{Answer Set Synthesis}

The conventional view of the ASP paradigm is in terms of problem solving.
The problem is described using \AnsProlog in such a way that the answer sets 
can be interpreted as the solutions to the problem. So the program provides the
specification for a solution; the constraints that elements in the search space have to satisfy.
For \systemname{} we use ASP for synthesis rather than problem
solving.  We are doing something subtly, but importantly
different; we are using a solver to generate representative objects
given a specification.  This is more knowledge representation than
problem solving, since the language is used to provide a computational
description of the required object.  We believe that this approach,
\emph{Answer Set Synthesis} opens new of possibilities for the use of \AnsProlog as there are lots of applications for which
we need parametrisable, consistent but not necessarily exceptional content.
Take for example, the automatic generation of a virtual world: one specifies the components of the environment and
their dependencies and the program will generate you the possible
locations. Generation of test data and puzzles are just a few more examples.

\subsection{ASP Methodology}

In constructing \systemname{ } a number of advantages of using answer set
programming have become clear, as have a number of limitations.

Firstly, \AnsProlog programs are very fast to write and very compact.  As well
as the obvious benefits, this means it is possible to develop the
system at the same time as undertaking knowledge capture and to
prototype features in the light of the advice of domain experts.  Part
of the reason why it is so fast to use is that rules are fully
declarative.  Programming thus focuses on expressing the concepts that
are being modelled rather than having to worry about which order to
put things in --- such as which rules should come first, which concepts have
higher priority, which choices should be made first, etc.  This also
makes incremental development easy as new constraints can be added one
at a time, without having to consider how they affect the search
strategy.

Being able to add rules incrementally during development turns out to
be extremely useful \cite{bcdv09} from a software engineering view point.  During
the development of \systemname, we experimented with a number of
different development methodologies.  As argued in \cite{cdvbp08}, 
``visualisation'' of answer sets is a very productive way bridging the 
gap between the problem domain and the program.
For the musical application domain, the most effective approach was
found to be first writing a script that translates answer sets to
human readable score or output for a synthesiser.  
Next the choice rules were added to the program to create all possible
pieces, valid or not.  Finally the constraints were incrementally
added to restrict the output to only valid sequences.  By building up
a library of valid pieces it was possible to perform regression
testing at each step and thus isolate bugs as soon as they were
introduced.

Using answer set programming was not without issue.  One persistent
problem was the lack of mature development support tools, particularly
debugging tools.  SPOCK \cite{brgepusctowo07b} was used but as its
focus is on computing the reasons behind the error, rather than the
interface issues of explaining these reasons to the user, it was
normally quicker to find bugs by looking at the last changes made and
which regression tests failed.  Generally, the bugs that where
encountered were due to subtle mismatches between the intended
meaning of a rule and the declarative reading of the rule used.  For
example the predicate \texttt{stepUp(P,T)} is used to represent the
proposition ``At time T, part P steps up to give the note at time
T+1'', however, it could easily be misinterpreted as ``At time T-1,
part P steps up to give the note at time T''. Which of these is used
is not important, as long as the same declarative reading is used consistently 
for all rules.  With the first ``meaning'' selected for \systemname, the
rule:\\

{\small
\begin{ttfamily}\begin{tabular}{ll}
chosenNote(1,T,N+S) :- & chosenNote(P,T-1,N), stepUp(P,T), \\
& stepBy(P,T,S).
\end{tabular}
\end{ttfamily}}

\noindent would not encode the intended progression of notes.
Even for experienced programmers, maintaining consistency in the semantics of predicates
can be hard. 
One possible way of supporting a programmer in avoiding these subtle
errors would be to develop a system that translated rules into natural
language, given the declarative reading of the propositions involved.
It should then be relatively straightforward to check that the rule
encoded what was intended.

\section{Conclusions and Future Work}\label{future}

We have presented an algorithmic composition system that uses \AnsProlog to
describe rules captured from classical texts on first species 
counterpoint.  The early system of \citeN{bbdf} has been developed in
two significant ways: greatly improved performance and the
introduction of a coherent rhythm schema.  However \systemname{} still has considerable scope for further development and enhancement.

The encoding of rhythm is still an area of active research. Although we now have a basic encoding
of rhythm it still requires a lot of fine tuning. The use of Farey
sequences and partitioning trees is the core
of this. In Section~\ref{rhythm} we already indicated that we intend to encode Sima Arom's principles which
he discovered when studying African Polyrhythms. The use of partitioning trees will need to be further developed
to do so.
 
We have concentrated on first species counterpoint from an early time, but many
of the rules apply to other styles.  We have made a few experiments
with rules for Bach chorales and for hymn tunes.  We need to partition
the current rule-sets into building blocks to facilitate reuse.

The current system can write short melodies effectively and
efficiently.  Development work is still needed to take this to entire
pieces.  We can start from these melodic fragments but a longer piece
needs a variety of different harmonisations for the same melody, and
related melodies with the same harmonic structure and a number of
similar techniques.  We have not solved the difficult global structure
problem but it does create a starting point on which we can build a
system that is hierarchical over time scales; we have a mechanism for
building syntactically correct sentences, but these need to be built
into paragraph and chapters, as it were.
It is not clear if this will be achieved within the current ASP system, 
or by a procedural layer built on top if it, or some other scheme.

To make the system more user-friendly, there is a need for a user interface, 
probably graphical, to select 
from the options and styles.  We have avoided this so far, as the
Musician-Machine Interface is a separate specialist area, but there are plans
for such an interface to be designed in the next phase.

In real life pieces some of the rules are sometimes broken.  This could
be simulated by one of a number of extensions to answer set semantics
(preferences, consistency restoring rules, defensible
rules, etc.).  However how to systematize the knowledge of when it is
acceptable to break the rules and in which contexts it is `better' to
break them is an open problem.

For \systemname{ } we used \AnsProlog as a computational description language rather than
just a knowledge representation one. 
There is a tempting possibility to apply the same methodology and
approach to other areas of content, such as maybe game map
generation.  Initial experiments show promise.


\begin{thebibliography}{}

\bibitem[\protect\citeauthoryear{Allen}{Allen}{1983}]{Allen}
{\sc Allen, J.~F.} 1983.
\newblock {Maintaining Knowledge about Temporal Intervals}.
\newblock {\em CACM\/}~{\em 26}, 198--3.

\bibitem[\protect\citeauthoryear{Anders}{Anders}{2007}]{AndersPhD}
{\sc Anders, T.} 2007.
\newblock {Composing Music by Composing Rules: Design and Usage of a Generic
  Music Constraint System}.
\newblock Ph.D. thesis, Queen's University, Belfast, Department of Music.

\bibitem[\protect\citeauthoryear{Arom}{Arom}{1991}]{sa:apap}
{\sc Arom, S.} 1991.
\newblock {\em {African Polyphony and Polyrhythm}}.
\newblock Cambridge University Press, Cambridge.

\bibitem[\protect\citeauthoryear{Baral}{Baral}{2003}]{baralbook}
{\sc Baral, C.} 2003.
\newblock {\em {Knowledge Representation, Reasoning and Declarative Problem
  Solving}}.
\newblock Cambridge Press.

\bibitem[\protect\citeauthoryear{Baral and Gelfond}{Baral and
  Gelfond}{2000}]{baral2000}
{\sc Baral, C.} {\sc and} {\sc Gelfond, M.} 2000.
\newblock {Reasoning Agents in Dynamic Domains}.
\newblock In {\em Logic-based Artificial Intelligence}. Kluwer Academic
  Publishers, 257--279.

\bibitem[\protect\citeauthoryear{Bel}{Bel}{1998}]{Bol}
{\sc Bel, B.} 1998.
\newblock {Migrating Musical Concepts: An Overview of the Bol Processor}.
\newblock {\em Computer Music Journal\/}~{\em 22,\/}~2, 56--64.

\bibitem[\protect\citeauthoryear{Boenn}{Boenn}{2007}]{Farey}
{\sc Boenn, G.} 2007.
\newblock {Composing Rhythms Based Upon Farey Sequences}.
\newblock In {\em Digital Music Research Network Conference}.

\bibitem[\protect\citeauthoryear{Boenn}{Boenn}{2008}]{gb:husserl}
{\sc Boenn, G.} 2008.
\newblock {The Importance of Husserl's Phenomenology of Internal
  Time-Consciousness for Music Analysis and Composition}.
\newblock In {\em Proceedings of the ICMC 2008}. Belfast.

\bibitem[\protect\citeauthoryear{Boenn}{Boenn}{2009}]{Boenn09}
{\sc Boenn, G.} 2009.
\newblock {Automated Analysis and Transcription of Rhythm Data and their Use
  for Composition}.
\newblock Ph.D. thesis, University of Bath.
\newblock submitted.

\bibitem[\protect\citeauthoryear{Boenn, Brain, De~Vos, and ffitch}{Boenn
  et~al\mbox{.}}{2008}]{bbdf}
{\sc Boenn, G.}, {\sc Brain, M.}, {\sc De~Vos, M.}, {\sc and} {\sc ffitch, J.}
  2008.
\newblock {Automatic Composition of Melodic and Harmonic Music by Answer Set
  Programming}.
\newblock In {\em International Conference on Logic Programming, ICLP08}.
  Lecture Notes in Computer Science, vol. 4386. Springer Berlin / Heidelberg,
  160--174.
\newblock in print.

\bibitem[\protect\citeauthoryear{Boulanger}{Boulanger}{2000}]{csound}
{\sc Boulanger, R.}, Ed. 2000.
\newblock {\em {The Csound Book: Tutorials in Software Synthesis and Sound
  Design}}.
\newblock MIT Press.

\bibitem[\protect\citeauthoryear{Brain, Cliffe, and De~Vos}{Brain
  et~al\mbox{.}}{2009}]{bcdv09}
{\sc Brain, M.}, {\sc Cliffe, O.}, {\sc and} {\sc De~Vos, M.} 2009.
\newblock {A Pragmatic Programmer's Guide to Answer Set Programming}.
\newblock In {\em Proceedings of the 2nd Workshop of Software Engineering for
  Answer Set Programming (SEA'09)}. 49--63.

\bibitem[\protect\citeauthoryear{Brain, Crick, {De Vos}, and Fitch}{Brain
  et~al\mbox{.}}{2006}]{bcdvf06b}
{\sc Brain, M.}, {\sc Crick, T.}, {\sc {De Vos}, M.}, {\sc and} {\sc Fitch, J.}
  2006.
\newblock {{TOAST}: Applying Answer Set Programming to Superoptimisation}.
\newblock In {\em International Conference on Logic Programming}. LNCS.
  Springer.

\bibitem[\protect\citeauthoryear{Brain, {De Vos}, and Satoh}{Brain
  et~al\mbox{.}}{2007}]{smodels-ie}
{\sc Brain, M.}, {\sc {De Vos}, M.}, {\sc and} {\sc Satoh, K.} 2007.
\newblock {\textsc{Smodels-IE}: Improving the cache Utilisation of Smodels}.
\newblock In {\em Proceedings of the 4th Workshop on Answer Set Programming:
  Advances in Theory and Implementation}, {S.~Costantini} {and} {R.~Watson},
  Eds. Porto, Portugal, 309--313.

\bibitem[\protect\citeauthoryear{Brain, Gebser, P{\"u}hrer, Schaub, Tompits,
  and Woltran}{Brain et~al\mbox{.}}{2007}]{brgepusctowo07b}
{\sc Brain, M.}, {\sc Gebser, M.}, {\sc P{\"u}hrer, J.}, {\sc Schaub, T.}, {\sc
  Tompits, H.}, {\sc and} {\sc Woltran, S.} 2007.
\newblock {{``}{T}hat is Illogical Captain!{"} -- {T}he Debugging Support Tool
  spock for Answer-Set Programs: System Description}.
\newblock In {\em Proceedings of the Workshop on Software Engineering for
  Answer Set Programming (SEA'07)}, {M.~{De Vos}} {and} {T.~Schaub}, Eds.
  71--85.

\bibitem[\protect\citeauthoryear{Brothwell and ffitch}{Brothwell and
  ffitch}{2008}]{Brothwell}
{\sc Brothwell, A.} {\sc and} {\sc ffitch, J.} 2008.
\newblock {An Automatic Blues Band}.
\newblock In {\em 6th International Linux Audio Conference}, {F.~Barknecht}
  {and} {M.~Rumori}, Eds. Tribun EU, Gorkeho 41, Bruno 602 00, Kunsthochscule
  f\"ur Medien K\"oln, 12--17.

\bibitem[\protect\citeauthoryear{Buccafurri and Caminiti}{Buccafurri and
  Caminiti}{2005}]{buccafurri+caminiti:2005}
{\sc Buccafurri, F.} {\sc and} {\sc Caminiti, G.} 2005.
\newblock {A Social Semantics for Multi-agent Systems}.
\newblock In {\em LPNMR}, {C.~Baral}, {G.~Greco}, {N.~Leone}, {and}
  {G.~Terracina}, Eds. Lecture Notes in Computer Science, vol. 3662. Springer,
  317--329.

\bibitem[\protect\citeauthoryear{Buccafurri and Gottlob}{Buccafurri and
  Gottlob}{2002}]{bucca2002}
{\sc Buccafurri, F.} {\sc and} {\sc Gottlob, G.} 2002.
\newblock {Multiagent Compromises, Joint Fixpoints, and Stable Models}.
\newblock In {\em Computational Logic: Logic Programming and Beyond, Essays in
  Honour of Robert A. Kowalski, Part I}, {A.~C. Kakas} {and} {F.~Sadri}, Eds.
  Lecture {N}otes in {C}omputer {S}cience, vol. 2407. Springer, 561--585.

\bibitem[\protect\citeauthoryear{Chuang}{Chuang}{1995}]{Dicegame}
{\sc Chuang, J.} 1995.
\newblock {Mozart's Musikalisches W\"urfelspiel}.
\newblock \url{http://sunsite.univie.ac.at/Mozart/dice/}.

\bibitem[\protect\citeauthoryear{Cliffe, De~Vos, Brain, and Padget}{Cliffe
  et~al\mbox{.}}{2008}]{cdvbp08}
{\sc Cliffe, O.}, {\sc De~Vos, M.}, {\sc Brain, M.}, {\sc and} {\sc Padget, J.}
  2008.
\newblock {{ASPVIZ}: Declarative Visualisation and Animation Using Answer Set
  Programming}.
\newblock In {\em Logic Programming}. Lecture Notes in Computer Science.
  Springer Berlin / Heidelberg, 724--728.

\bibitem[\protect\citeauthoryear{Cliffe, {De Vos}, and Padget}{Cliffe
  et~al\mbox{.}}{2006}]{cdvp06c}
{\sc Cliffe, O.}, {\sc {De Vos}, M.}, {\sc and} {\sc Padget, J.} 2006.
\newblock {Specifying and Analysing Agent-based Social Institutions using
  Answer Set Programming}.
\newblock In {\em Selected revised papers from the workshops on Agent, Norms
  and Institutions for Regulated Multi-Agent Systems (ANIREM) and Organizations
  and Organization Oriented Programming (OOOP) at AAMAS'05}, {O.~Boissier},
  {J.~Padget}, {V.~Dignum}, {G.~Lindemann}, {E.~Matson}, {S.~Ossowski},
  {J.~Sichman}, {and} {J.~Vazquez-Salceda}, Eds. LNCS, vol. 3913. Springer
  Verlag, 99--113.

\bibitem[\protect\citeauthoryear{Cope}{Cope}{2006}]{Cope}
{\sc Cope, D.} 2006.
\newblock {A Musical Learning Algorithm}.
\newblock {\em Computer Music Journal\/}~{\em 28,\/}~3 (Fall), 12--27.

\bibitem[\protect\citeauthoryear{de~la Motte}{de~la Motte}{1981}]{ddlm:kp}
{\sc de~la Motte, D.} 1981.
\newblock {\em {Kontrapunkt}}.
\newblock dtv/B{\"a}renreiter, M{\"u}nchen.

\bibitem[\protect\citeauthoryear{De~Vos and Vermeir}{De~Vos and
  Vermeir}{1999}]{dvver99a}
{\sc De~Vos, M.} {\sc and} {\sc Vermeir, D.} 1999.
\newblock {C}hoice {L}ogic {P}rograms and {N}ash {E}quilibria in {S}trategic
  {G}ames.
\newblock In {\em Computer Science Logic (CSL'99)}, {J.~Flum} {and}
  {M.~Rodr{\'\i}guez-Artalejo}, Eds. Lecture Notes in Computer Science, vol.
  1683. Springer Verslag, Madrid, Spain, 266--276.

\bibitem[\protect\citeauthoryear{De~Vos and Vermeir}{De~Vos and
  Vermeir}{2004}]{dvver04}
{\sc De~Vos, M.} {\sc and} {\sc Vermeir, D.} 2004.
\newblock Extending {A}nswer {S}ets for {L}ogic {P}rogramming {A}gents.
\newblock {\em Annals of Mathematics and Artifical Intelligence\/}~{\em
  42,\/}~1--3 (Sept.), 103--139.
\newblock Special Issue on Computational Logic in Multi-Agent Systems.

\bibitem[\protect\citeauthoryear{Ebcio\u{g}lu}{Ebcio\u{g}lu}{1986}]{Ebcioglu}
{\sc Ebcio\u{g}lu, K.} 1986.
\newblock {An Expert System for Harmonization of Chorales in the Style of J.S.
  Bach}.
\newblock Ph.D. thesis, State University of New York, Buffalo, Department of
  Computer Science.

\bibitem[\protect\citeauthoryear{Eiter, Faber, Leone, and Pfeifer}{Eiter
  et~al\mbox{.}}{1999}]{eiter99}
{\sc Eiter, T.}, {\sc Faber, W.}, {\sc Leone, N.}, {\sc and} {\sc Pfeifer, G.}
  1999.
\newblock {The Diagnosis Frontend of the {DLV} System}.
\newblock {\em AI Communications\/}~{\em 12,\/}~1-2, 99--111.

\bibitem[\protect\citeauthoryear{Eiter, Faber, Leone, Pfeifer, and
  Polleres}{Eiter et~al\mbox{.}}{2002}]{eiter2002}
{\sc Eiter, T.}, {\sc Faber, W.}, {\sc Leone, N.}, {\sc Pfeifer, G.}, {\sc and}
  {\sc Polleres, A.} 2002.
\newblock {The {DLV}$^K$ Planning System}.
\newblock In {\em European Conference, JELIA 2002}, {S.~Flesca}, {S.~Greco},
  {N.~Leone}, {and} {G.~Ianni}, Eds. LNAI, vol. 2424. Springer Verlag, Cosenza,
  Italy, 541--544.

\bibitem[\protect\citeauthoryear{Eiter, Leone, Mateis, Pfeifer, and
  Scarcello}{Eiter et~al\mbox{.}}{1998}]{eilemapfsc98}
{\sc Eiter, T.}, {\sc Leone, N.}, {\sc Mateis, C.}, {\sc Pfeifer, G.}, {\sc
  and} {\sc Scarcello, F.} 1998.
\newblock {The {KR} System {\tt DLV}: Progress Report, Comparisons and
  Benchmarks}.
\newblock In {\em {KR}'98: Principles of Knowledge Representation and
  Reasoning}, {A.~G. Cohn}, {L.~Schubert}, {and} {S.~C. Shapiro}, Eds. Morgan
  Kaufmann, San Francisco, California, 406--417.

\bibitem[\protect\citeauthoryear{Erdem, Lifschitz, Nakhleh, and Ringe}{Erdem
  et~al\mbox{.}}{2003}]{ErdemLNR03}
{\sc Erdem, E.}, {\sc Lifschitz, V.}, {\sc Nakhleh, L.}, {\sc and} {\sc Ringe,
  D.} 2003.
\newblock {Reconstructing the Evolutionary History of Indo-European Languages
  Using Answer Set Programming}.
\newblock In {\em PADL}, {V.~Dahl} {and} {P.~Wadler}, Eds. LNCS, vol. 2562.
  Springer, 160--176.

\bibitem[\protect\citeauthoryear{Erdem, Lifschitz, and Ringe}{Erdem
  et~al\mbox{.}}{2006}]{ErdemLR06}
{\sc Erdem, E.}, {\sc Lifschitz, V.}, {\sc and} {\sc Ringe, D.} 2006.
\newblock {Temporal Phylogenetic Networks and Logic Programming}.
\newblock {\em TPLP\/}~{\em 6,\/}~5, 539--558.

\bibitem[\protect\citeauthoryear{Fux}{Fux}{1725}]{Fux}
{\sc Fux, J.} 1965, orig 1725.
\newblock {\em {The Study of Counterpoint from Johann Joseph Fux's Gradus ad
  Parnassum}}.
\newblock W.W. Norton.

\bibitem[\protect\citeauthoryear{Gebser, Kaufmann, Neumann, and Schaub}{Gebser
  et~al\mbox{.}}{2007}]{gekanesc07a}
{\sc Gebser, M.}, {\sc Kaufmann, B.}, {\sc Neumann, A.}, {\sc and} {\sc Schaub,
  T.} 2007.
\newblock {Conflict-Driven Answer Set Solving}.
\newblock In {\em Proceedings of the Twentieth International Joint Conference
  on Artificial Intelligence (IJCAI'07)}, {M.~Veloso}, Ed. AAAI Press/The MIT
  Press, 386--392.
\newblock Available at http://www.ijcai.org/papers07/contents.php.

\bibitem[\protect\citeauthoryear{Gebser, Schaub, and Thiele}{Gebser
  et~al\mbox{.}}{2007}]{gescth07a}
{\sc Gebser, M.}, {\sc Schaub, T.}, {\sc and} {\sc Thiele, S.} 2007.
\newblock {GrinGo: A New Grounder for Answer Set Programming}.
\newblock In {\em Proceedings of the Ninth International Conference on Logic
  Programming and Nonmonotonic Reasoning (LPNMR'07)}, {C.~Baral}, {G.~Brewka},
  {and} {J.~Schlipf}, Eds. Lecture Notes in Artificial Intelligence, vol. 4483.
  Springer-Verlag, 266--271.

\bibitem[\protect\citeauthoryear{Gelfond and Lifschitz}{Gelfond and
  Lifschitz}{1988}]{gellif88}
{\sc Gelfond, M.} {\sc and} {\sc Lifschitz, V.} 1988.
\newblock {The Stable Model Semantics for Logic Programming}.
\newblock In {\em Logic Programming, Proceedings of the Fifth International
  Conference and Symposium}, {R.~A. Kowalski} {and} {K.~A. Bowen}, Eds. The MIT
  Press, Seattle, Washington, 1070--1080.

\bibitem[\protect\citeauthoryear{Gelfond and Lifschitz}{Gelfond and
  Lifschitz}{1991}]{gellif91}
{\sc Gelfond, M.} {\sc and} {\sc Lifschitz, V.} 1991.
\newblock {Classical Negation in Logic Programs and Disjunctive Databases}.
\newblock {\em New Generation Computing\/}~{\em 9,\/}~3-4, 365--386.

\bibitem[\protect\citeauthoryear{Giunchiglia, Lierler, and Maratea}{Giunchiglia
  et~al\mbox{.}}{2004}]{sat-asp}
{\sc Giunchiglia, E.}, {\sc Lierler, Y.}, {\sc and} {\sc Maratea, M.} 2004.
\newblock {{SAT-Based Answer Set Programming}}.
\newblock In {\em {Proceedings of the 18th National Conference on Artificial
  Intelligence (AAAI-04)}}. 61--66.

\bibitem[\protect\citeauthoryear{Goranko, Montanari, and Sciavicco}{Goranko
  et~al\mbox{.}}{2003}]{Goranko}
{\sc Goranko, V.}, {\sc Montanari, A.}, {\sc and} {\sc Sciavicco, G.} 2003.
\newblock {A Road Map on Interval Temporal Logics and Duration Calculi}.
\newblock {\em Journal of Applied Non-Classical Logics\/}~{\em 14}, 9--54.

\bibitem[\protect\citeauthoryear{Grell, Schaub, and Selbig}{Grell
  et~al\mbox{.}}{2006}]{grscse06b}
{\sc Grell, S.}, {\sc Schaub, T.}, {\sc and} {\sc Selbig, J.} 2006.
\newblock {{Modelling Biological Networks by Action Languages via Answer Set
  Programming}}.
\newblock In {\em Proceedings of the International Conference on Logic
  Programming (ICLP'06)}, {S.~Etalle} {and} {M.~Truszczy\'nski}, Eds. LNCS,
  vol. 4079. Springer-Verlag, 285--299.

\bibitem[\protect\citeauthoryear{Gressmann, Janhunen, Mercer, Schaub, Thiele,
  and Tichy}{Gressmann et~al\mbox{.}}{2005}]{grjamescthti05a}
{\sc Gressmann, J.}, {\sc Janhunen, T.}, {\sc Mercer, R.}, {\sc Schaub, T.},
  {\sc Thiele, S.}, {\sc and} {\sc Tichy, R.} 2005.
\newblock {{Platypus: A Platform for Distributed Answer Set Solving}}.
\newblock In {\em Proceedings of the 8th International Conference on Logic
  Programming and Nonmonotonic Reasoning (LPNMR'05)}. 227--239.

\bibitem[\protect\citeauthoryear{Hajdu}{Hajdu}{1993}]{hajdu}
{\sc Hajdu, G.} 1993.
\newblock {Low Energy and Equal Spacing. The Multifactorial Evolution of Tuning
  Systems}.
\newblock {\em Interface\/}~{\em 22}, 319--333.

\bibitem[\protect\citeauthoryear{Hardy and Wright}{Hardy and
  Wright}{1938}]{hardy}
{\sc Hardy, G.} {\sc and} {\sc Wright, E.} 1938.
\newblock {\em {An Introduction to the Theory of Numbers}\/}, 4th ed.
\newblock Oxford University Press.

\bibitem[\protect\citeauthoryear{Konczak}{Konczak}{2006}]{konczak06b}
{\sc Konczak, K.} 2006.
\newblock {{Voting Theory in Answer Set Programming}}.
\newblock In {\em Proceedings of the Twentieth Workshop on Logic Programming
  (WLP'06)}, {M.~Fink}, {H.~Tompits}, {and} {S.~Woltran}, Eds. Number INFSYS
  RR-1843-06-02 in Technical Report Series. Technische Universit{\"a}t Wien,
  45--53.

\bibitem[\protect\citeauthoryear{Leach}{Leach}{1999}]{Leach}
{\sc Leach, J.~L.} 1999.
\newblock {Algorithmic Composition and Musical Form}.
\newblock Ph.D. thesis, University of Bath, School of Mathematical Sciences.

\bibitem[\protect\citeauthoryear{Lewis}{Lewis}{2000}]{Lewis}
{\sc Lewis, G.~E.} 2000.
\newblock {Too Many Notes: Computers, Complexity and Culture in Voyager}.
\newblock {\em Leonardo Music Journal\/}~{\em 10}, 33--39.

\bibitem[\protect\citeauthoryear{Lierler}{Lierler}{2008}]{sup}
{\sc Lierler, Y.} 2008.
\newblock {Abstract Answer Set Solvers}.
\newblock In {\em ICLP '08: Proceedings of the 24th International Conference on
  Logic Programming}. Springer-Verlag, Berlin, Heidelberg, 377--391.

\bibitem[\protect\citeauthoryear{Lierler and Maratea}{Lierler and
  Maratea}{2004}]{cmodels}
{\sc Lierler, Y.} {\sc and} {\sc Maratea, M.} 2004.
\newblock {Cmodels-2: SAT-based Answer Set Solver Enhanced to Non-tight
  Programs}.
\newblock In {\em Proceedings of the 7th International Conference on Logic
  Programming and Nonmonotonic Reasoning}. LNCS, vol. 2923. Springer, 346--350.

\bibitem[\protect\citeauthoryear{Lifschitz}{Lifschitz}{2002}]{lif2000}
{\sc Lifschitz, V.} 2002.
\newblock {Answer Set Programming and Plan Generation}.
\newblock {\em J. of Artificial Intelligence\/}~{\em 138,\/}~1-2, 39--54.

\bibitem[\protect\citeauthoryear{Mileo and Schaub}{Mileo and
  Schaub}{2006}]{milsch06a}
{\sc Mileo, A.} {\sc and} {\sc Schaub, T.} 2006.
\newblock {{Extending Ordered Disjunctions for Policy Enforcement: Preliminary
  report}}.
\newblock In {\em Proceedings of the International Workshop on Preferences in
  Logic Programming Systems (PREFS'06)}, {E.~Pontelli} {and} {T.~Son}, Eds.
  45--59.

\bibitem[\protect\citeauthoryear{Niemel{\"a} and Simons}{Niemel{\"a} and
  Simons}{1997}]{nisi97}
{\sc Niemel{\"a}, I.} {\sc and} {\sc Simons, P.} 1997.
\newblock {Smodels: An Implementation of the Stable Model and Well-founded
  Semantics for Normal {LP}}.
\newblock In {\em Proceedings of the 4th International Conference on Logic
  Programing and Nonmonotonic Reasoning}, {J.~Dix}, {U.~Furbach}, {and}
  {A.~Nerode}, Eds. LNAI, vol. 1265. Springer, Berlin, 420--429.

\bibitem[\protect\citeauthoryear{Niemel{\"a}, Simons, and Soininen}{Niemel{\"a}
  et~al\mbox{.}}{1999}]{NiemelaSS99}
{\sc Niemel{\"a}, I.}, {\sc Simons, P.}, {\sc and} {\sc Soininen, T.} 1999.
\newblock {Stable Model Semantics of Weight Constraint Rules}.
\newblock In {\em LPNMR}, {M.~Gelfond}, {N.~Leone}, {and} {G.~Pfeifer}, Eds.
  Lecture Notes in Computer Science, vol. 1730. Springer, 317--331.

\bibitem[\protect\citeauthoryear{Padovani and Provetti}{Padovani and
  Provetti}{2004}]{PadovaniP04}
{\sc Padovani, L.} {\sc and} {\sc Provetti, A.} 2004.
\newblock {Qsmodels: {ASP} Planning in Interactive Gaming Environment}.
\newblock In {\em JELIA}, {J.~J. Alferes} {and} {J.~A. Leite}, Eds. Lecture
  Notes in Computer Science, vol. 3229. Springer, 689--692.

\bibitem[\protect\citeauthoryear{Partch}{Partch}{1979}]{partch1979}
{\sc Partch, H.} 1979.
\newblock {\em {Genesis of a Music}}.
\newblock Da Capo Press, New York.

\bibitem[\protect\citeauthoryear{Rohrmeier}{Rohrmeier}{2006}]{Rohrmeier}
{\sc Rohrmeier, M.} 2006.
\newblock {Towards Modelling Harmonic Movement in Music: Analysing Properties
  and Dynamic Aspects of pc set Sequences in Bach's Chorales}.
\newblock Tech. Rep. DCRR-004, Darwin College, University of Cambridge.

\bibitem[\protect\citeauthoryear{Soininen and Niemel\"a}{Soininen and
  Niemel\"a}{1999}]{soininen99}
{\sc Soininen, T.} {\sc and} {\sc Niemel\"a, I.} 1999.
\newblock {Developing a Declarative Rule Language for Applications in Product
  Configuration}.
\newblock In {\em Proceedings of the First International Workshop on Practical
  Aspects of Declarative Languages (PADL '99)}. LNCS. Springer, San Antonio,
  Texas.

\bibitem[\protect\citeauthoryear{Syrj{\"a}nen and Niemel{\"a}}{Syrj{\"a}nen and
  Niemel{\"a}}{2001}]{smodels}
{\sc Syrj{\"a}nen, T.} {\sc and} {\sc Niemel{\"a}, I.} 2001.
\newblock {The Smodels System}.
\newblock In {\em Proceedings of the 6th International Conference on Logic
  Programming and Nonmonotonic Reasoning}.

\bibitem[\protect\citeauthoryear{Thakar}{Thakar}{1990}]{Thakar}
{\sc Thakar, M.} 1990.
\newblock {\em {Counterpoint}}.
\newblock New Haven.

\bibitem[\protect\citeauthoryear{Ward and Schlipf}{Ward and
  Schlipf}{2004}]{smodelscc}
{\sc Ward, J.} {\sc and} {\sc Schlipf, S.} 2004.
\newblock {Answer Set Programming with Clause Learning}.
\newblock In {\em Proceedings of the 7th International Conference on Logic
  Programming and Nonmonotonic Reasoning}. LNCS, vol. 2923. Springer.

\bibitem[\protect\citeauthoryear{Xenakis}{Xenakis}{1992}]{Xenakis}
{\sc Xenakis, I.} 1992.
\newblock {\em {Formalized Music}}.
\newblock Bloomington Press, Stuyvesant, NY, USA.

\end{thebibliography}

\end{document}